\documentclass{article}
\usepackage{graphicx}
\usepackage{hyperref}
\usepackage{lineno}
\usepackage[OT1]{fontenc}
\usepackage{amsthm,amsmath,amssymb}
\usepackage[numbers]{natbib}
\usepackage{graphicx}
\usepackage{subfig}
\usepackage{titlesec}
\usepackage{mathrsfs}
\usepackage{multirow}
\usepackage{booktabs}
\RequirePackage{enumitem} 
\RequirePackage{mathtools}
\usepackage[doublespacing]{setspace}
\usepackage{geometry}
\usepackage{subfig}
\usepackage[toc]{appendix}
\usepackage{color}
\allowdisplaybreaks

\numberwithin{equation}{section}
\theoremstyle{plain}

\titlelabel{\thetitle.\quad}

\makeatletter
\renewcommand\@biblabel[1]{} % No brackets for the references
\renewenvironment{thebibliography}[1]
{\section*{\refname}%
	\@mkboth{\MakeUppercase\refname}{\MakeUppercase\refname}%
	\list{\@biblabel{\@arabic\c@enumiv}}%
	{\settowidth\labelwidth{\@biblabel{}}%
		\leftmargin\labelwidth
		\advance\leftmargin15pt% change 20 pt according to your needs
		\advance\leftmargin\labelsep
		\setlength\itemindent{-10pt}% change using the inverse of the length used before
		\@openbib@code
		\usecounter{enumiv}%
		\let\p@enumiv\@empty
		\renewcommand\theenumiv{\@arabic\c@enumiv}}%
	\sloppy
	\clubpenalty4000
	\@clubpenalty \clubpenalty
	\widowpenalty4000%
	\sfcode`\.\@m}
{\def\@noitemerr
	{\@latex@warning{Empty `thebibliography' environment}}%
	\endlist}
\renewcommand\newblock{\hskip .11em\@plus.33em\@minus.07em}
\makeatother

\renewcommand{\refname}{References}
\begin{document}

\title{\textbf{Blinded and unblinded sample size re-estimation in crossover trials balanced for period}}
\author{\textbf{M. J. Grayling\textsuperscript{1}, A. P. Mander\textsuperscript{1}, J. M. S. Wason\textsuperscript{1,2}}\\
1. Hub for Trials Methodology Research, MRC Biostatistics Unit, Cambridge, UK, \\ 2. Institute of Health and Society, Newcastle University, Newcastle, UK.}
\date{}
\maketitle

\noindent \textbf{Corresponding Author:} M. J. Grayling, MRC Biostatistics Unit, Forvie Site, Robinson Way, Cambridge CB2 0SR, UK; Tel: +44-(0)1223-330300; E-mail: mjg211@cam.ac.uk.\\

\noindent \textbf{Abstract:} The determination of the sample size required by a crossover trial typically depends on the specification of one or more variance components. Uncertainty about the value of these parameters at the design stage means that there is often a risk a trial may be under- or over-powered. For many study designs, this problem has been addressed by considering adaptive design methodology that allows for the re-estimation of the required sample size during a trial. Here, we propose and compare several approaches for this in multi-treatment crossover trials. Specifically, regulators favour re-estimation procedures to maintain the blinding of the treatment allocations. We therefore develop blinded estimators for the within and between person variances, following simple or block randomisation. We demonstrate that, provided an equal number of patients are allocated to sequences that are balanced for period, the proposed estimators following block randomisation are unbiased. We further provide a formula for the bias of the estimators following simple randomisation. The performance of these procedures, along with that of an unblinded approach, is then examined utilising three motivating examples, including one based on a recently completed four-treatment four-period crossover trial. Simulation results show that the performance of the proposed blinded procedures is in many cases similar to that of the unblinded approach, and thus they are an attractive alternative.
\\

\\
\section{Introduction}

Crossover trials, in which participants are randomly allocated to receive a sequence of treatments across a series of time periods, are an extremely useful tool in clinical research. Their nature permits each patient to act as their own control, exploiting the fact that in most instances the variability of measurements on different subjects in a study will be far greater than that on the same subject. In this way, crossover trials are often more efficient than parallel group trials. Like most experimental designs, the determination of the sample size required by a crossover trial, to achieve a certain power for a particular treatment effect, depends on the significance level, and at least one factor which accounts for the participant's variance in response to treatment. Whilst the former are designated quantities, the variance factors will usually be subject to substantial uncertainty at the design stage. Their value will often be greatly affected by components of the current trial, such as inclusion/exclusion criteria for example, that renders estimates obtained from previous trials biased. This is troubling since sample size calculation is of paramount importance in study design. Planning a trial that is too large results in an unnecessary number of patients being made susceptible to interventions that may be harmful. It also needlessly wastes valuable resources in terms of time, money, and available trial participants. In contrast, too small a sample size confers little chance of success for a trial. The consequences of this could be far reaching: a wrong decision may lead to the halting of the development of a therapy, which could deprive future patients of a valuable treatment option.

To address this problem in a parallel group setting with normally distributed outcome variables, \citet{wittes1990}, building upon previous work by \citet{stein1945}, proposed the internal pilot study design. In their approach, at an interim time period the accrued data is unblinded, the within-group variance computed, and the trial's required sample size adjusted if necessary. However, unblinding an ongoing trial can reduce its integrity and introduce bias \citep{ich1998}. Consequently, \citet{gould1992} explored several approaches for re-estimating the required sample size in a blinded manner. Since then, a number of papers have advocated for re-estimation in a parallel group setting to be based upon a crude one-sample estimate of the variance, and methodology has also been proposed which allows the type-I error-rate to be more accurately controlled \citep{kieser2003}. More recently, much work has been conducted on similar methods for an array of possible trial designs and types of outcome variable (see, e.g., \citet{jensen2010} and \citet{togo2011}), with these methods also gaining regulatory acceptance \citep{chmp2007,fda2010}.

Thus, today, sample size re-estimation procedures have established themselves for parallel group trials as an advantageous method to employ when there is pre-trial uncertainty over the appropriate sample size. In contrast, there has been little exploration of such methodology within the context of multi-treatment crossover trials. \citet{golkowski2014} recently explored a blinded sample size re-estimation procedure for establishing bioequivalence in a trial utilising an AB/BA crossover design. \citet{jones2014} discussed how the results of \citet{kieser2003} could be rephrased for an AB/BA crossover trial testing for superiority. In addition, several unblinded re-estimation procedures for AB/BA bioequivalence trials have been proposed \citep{potvin2007,montague2012,xu2016}, the performance of which has recently been extensively compared \citep{kieser2015}. The work of \citet{lake2002} and \citet{schie2014} on sample size re-estimation in cluster randomised trials has some parallels with the methodology required for crossover trials, because of the necessitated mixed model for data analysis. Likewise, this is true of the methodology presented by \citet{zucker2002} on re-estimation procedures for longitudinal trials. However, we are unaware of any article that explicitly discusses re-estimation in crossover trials with more than two-treatments. There are many examples of such trials in the literature, whilst they also remain the focus of much research (see, e.g., \citet{bailey2014} and, \citet{lui2016}).

In this article, we consider several possible approaches to the interim re-assessment of the sample size required by a multi-treatment crossover trial. We assume a normally distributed outcome variable, and that a commonly utilised linear mixed model will be employed for data analysis. We focus primarily on a setting in which the final analysis is based on many-to-one comparisons for one-sided null hypotheses, but provide additional guidance for other possibilities in the Appendix. Blinded procedures for estimating the between and within person variance in response are proposed, following either simple or block randomisation to sequences that are balanced for period. The performance of these estimators is contrasted to that of an unblinded procedure via a simulation study motivated by a real four-treatment four-period crossover trial. Additionally, in the Appendix we provide results for two additional examples. We now proceed by specifying the notation used in the re-estimation procedures. Our findings are then summarised in Section 3, before we conclude in Section 4 with a discussion.

\section{Methods}

\subsection{Hypotheses, notation, and analysis}

We consider a crossover trial with $D$ treatments, indexed $d=0,\dots,D-1$. Treatments $d=1,...,D-1$ are considered experimental, and are to be compared to the common control $d=0$. We suppose that $K$ sequences, indexed $k=1,\dots,K$, are utilised for treatment allocation, and denote by $n_k$ the number of patients allocated to sequence $k$. The number of periods in the trial, which is equal to the length of each of the sequences, is denoted by $P$.

We restrict our focus to trials with normally distributed outcome data, to be analysed using the following linear mixed model
\begin{equation}\label{eq2}
y_{ijk} = \mu_0 + \pi_j + \tau_{\text{d}(j,k)} + s_{ik} + \epsilon_{ijk},\qquad i=1,\dots,n_k,\ j=1,\dots,P,\ k=1,\dots,K.
\end{equation}
Here
\begin{itemize}
	\item $y_{ijk}$ is the response for individual $i$, in period $j$, on sequence $k$;
	\item $\mu_0$ is an intercept term; the mean response on treatment 0 in period 1;
	\item $\pi_j$ is a fixed effect for period $j$, with the identifiability constraint $\pi_1=0$;
	\item $\tau_{\text{d}(j,k)}$ is a fixed direct treatment effect for the treatment administered to an individual in period $j$, on sequence $k$, with the identifiability constraint $\tau_0=0$. Thus $\text{d}(j,k)=0,\dots,D-1$;
	\item $s_{ik} \sim N(0,\sigma_b^2)$ is a random effect for individual $i$ on sequence $k$;
	\item $\epsilon_{ijk} \sim N(0,\sigma_e^2)$ is the residual for the response from individual $i$, in period $j$, on sequence $k$.
\end{itemize}
This model, and its implied covariance structure, is the standard for a crossover trial that ignores the possible effects of carryover. Thus we are implicitly heeding the advice of \citet{senn1992}, and others, that a crossover trial should not be conducted when carryover is likely to be an issue. Furthermore, note that by the above, two observations $y_{i_1j_1k_1}$ and $y_{i_2j_2k_2}$ are independent unless $i_1=i_2$ and $k_1=k_2$.

We assume that the following hypotheses are to be tested, to attempt to establish the superiority of each experimental intervention versus the control
\[ H_{0d} : \tau_d \le 0, \qquad H_{1d} : \tau_d > 0, \qquad d=1,\dots,D-1. \]
Note though that for Examples 1 and 3, slightly different hypotheses are assessed, as negative effects imply efficacy. Additionally, in the Appendix we detail how one can handle alternate hypotheses of interest.

We suppose that it is desired to strongly control the FWER, the maximal probability of one or more incorrect rejections amongst the family of null hypotheses for all possible treatment effects, to some specified level $\alpha\in(0,1)$. There are several possible ways to define power in a multi-treatment setting. Throughout, we assume that pairwise power of at least $1-\beta\in(0,1)$ to reject, without loss of generality, $H_{01}$ is required when $\tau_1=\delta>0$ for designated type-II error-rate $\beta$ and clinically relevant difference $\delta$. Thus, from here, when referring to power we mean the probability that $H_{01}$ is rejected. However, in the Appendix we describe how a desired familywise power could be achieved.

To test the hypotheses, we assume that $N$ patients in total will be recruited to the trial, with each randomised to one of the $K$ sequences, and that the the linear mixed model~(\ref{eq2}) will be fitted to the accumulated data. Note that in fitting this model, a choice must be made over whether to utilise maximum likelihood, or restricted error maximum likelihood (REML), estimation. Given the bias of the maximum likelihood estimator of the variance components of a linear mixed model in finite samples, and that crossover trials are often conducted with relatively small sample sizes, here we always take the latter approach. Note though that this would have little effect for larger sample sizes. For further details on these considerations, we refer the reader to, for example, \citet{fitzmaurice2011}. In brief, the REML estimation procedure, for a linear mixed model of the form $\boldsymbol{y}=\boldsymbol{X}\boldsymbol{\beta}+\boldsymbol{Z}\boldsymbol{b}+\boldsymbol{\epsilon}$ with $\boldsymbol{b}\sim N(\boldsymbol{0},\boldsymbol{G})$ and $\boldsymbol{\epsilon}\sim N(\boldsymbol{0},\boldsymbol{R})$, iteratively optimizes the parameter estimates for the effects in the model. The following modified log-likelihood is maximised to provide an estimate, $\hat{\boldsymbol{\Sigma}}$, for $\boldsymbol{\Sigma}=\boldsymbol{Z}\boldsymbol{G}\boldsymbol{Z}^\top+\boldsymbol{R}$, using an estimate, $\hat{\boldsymbol{\beta}}$, for $\boldsymbol{\beta}$
	\begin{equation*}
	-\frac{1}{2} \log{|\boldsymbol{\Sigma}|}-\frac{1}{2}(\boldsymbol{y}-\boldsymbol{X}\hat{\boldsymbol{\beta}})^\top\boldsymbol{\Sigma}^{-1}(\boldsymbol{y}-\boldsymbol{X}\hat{\boldsymbol{\beta}})-\frac{1}{2}\log{|\boldsymbol{X}^\top \boldsymbol{\Sigma}^{-1}\boldsymbol{X}|}.
	\end{equation*}
	Then, $\hat{\boldsymbol{\beta}}$ is updated to
	\begin{equation*}
	\hat{\boldsymbol{\beta}} = (\boldsymbol{X}^\top\hat{\boldsymbol{\Sigma}}^{-1}\boldsymbol{X})^{-1}\boldsymbol{X}^\top\hat{\boldsymbol{\Sigma}}^{-1}\boldsymbol{y},
	\end{equation*}
	and the process repeated. Given the final solutions $\hat{\boldsymbol{\beta}}$ and $\hat{\boldsymbol{\Sigma}}$, we take $\text{Var}(\hat{\boldsymbol{\beta}})=(\boldsymbol{X}^\top\hat{\boldsymbol{\Sigma}}^{-1}\boldsymbol{X})^{-1}$.
	
	In our case, $\boldsymbol{\beta}=(\mu_0,\pi_2,\dots,\pi_P,\tau_1,\dots,\tau_{D-1})^\top$, and the following $D-1$ Wald test statistics are formed
	\[ T_d = \frac{\hat{\tau}_d}{\sqrt{\text{Var}(\hat{\tau}_d)}}, \qquad d=1,\dots,D-1, \]
	where $\hat{\tau}_d$ and $\text{Var}(\hat{\tau}_d)$ are extracted from $\hat{\boldsymbol{\beta}}$ and $\text{Var}(\hat{\boldsymbol{\beta}})$ respectively.
	
	Next, we reject $H_{0d}$ if $T_d>e$, with $e$ chosen to control the FWER. Explicitly, using a Dunnett test \citep{dunnett1955}, we take $e$ as the solution to
	\begin{equation}\label{eq1}
	1 - \alpha = \Psi_{D-1}\{(e,\dots,e)^T,\text{Var}(\boldsymbol{T}),\nu_N\},
	\end{equation}
	where $\Psi_{M}\{\textbf{x},\Lambda,\nu\}$ is the $M$-dimensional cumulative distribution function of a central multivariate \textit{t}-distribution with covariance matrix $\Lambda$ and $\nu$ degrees of freedom. We take the degrees of freedom here, for sample size $N$, to be $\nu_N=(N-1)(P-1)-(D-1)$, which arises from that associated with an analogous multi-level ANOVA design. Moreover, $\text{Var}(\boldsymbol{T})$ is the covariance matrix of $\boldsymbol{T}=(T_1,\dots,T_{D-1})^\top$, which can be calculated using $\text{Var}(\hat{\boldsymbol{\beta}})$.
	
	Now, in this case, if $\sigma_e^2$ and $\sigma_b^2$ were known, and we assumed that $n_1=\dots=n_K$, we could derive a simple formula for the total number of patients, $N$, required to achieve the desired power for the trial. Here, we denote this formula using the function $\text{N}(\sigma_e^2,\sigma_b^2)$, explicitly stating it's dependence upon the within and between person variances. In the Appendix, we elaborate on how this formula can be derived.
	
	Our problem, as discussed, is that in practice $\sigma_e^2$ and $\sigma_b^2$ are rarely known accurately at the design stage. Therefore, we propose to re-estimate the required sample size at an interim analysis timed after $n_{\text{int}}\in\mathbb{N}$ patients. That is, we consider several methods to construct estimates, $\hat{\sigma}_e^2$ and $\hat{\sigma}_b^2$, for $\sigma_e^2$ and $\sigma_b^2$ respectively, based on the data accrued up to the interim analysis. Then, the final sample size for the trial is taken as
	
	\[\hat{N} = \left\{\begin{array}{ll}
	n_{\text{int}} & \text{if } \text{N}(\hat{\sigma}_e^2,\hat{\sigma}_b^2) \leq n_{\text{int}},\\
	\lceil \text{N}(\hat{\sigma}_e^2,\hat{\sigma}_b^2) \rceil & \text{if } n_{\text{int}} < \text{N}(\hat{\sigma}_e^2,\hat{\sigma}_b^2) < n_{\text{max}},\\
	n_{\text{max}} & \text{if } \text{N}(\hat{\sigma}_e^2,\hat{\sigma}_b^2) \geq n_{\text{max}},\end{array}\right.\]

where $\lceil x \rceil$ denotes the nearest integer greater than or equal to $x$ and $n_{\text{max}}\in\mathbb{N}$ is a specified maximal allowed sample size. It could be based, for example, on the cost restrictions or feasible recruitment rate of a trial. Of course, if $\hat{N} = n_{\text{max}}$ then the trial will be expected to be under-powered. Thus, if necessary, additional patients are recruited and a final analysis conducted as above based on the calculated values of the test statistics $T_d$, and the critical value $e$ as defined in Equation~(\ref{eq1}).

Throughout, to give our function $\text{N}(\cdot)$ a simple form, we consider values of $n_{\text{int}}$ that imply an equal number of patients could be allocated to each of the $K$ sequences, and assume randomisation schemes that ensures this is the case. Moreover, for reasons to be elucidated shortly, we consider from here only settings where the $K$ sequences are balanced for period. That is, across the chosen sequences, each treatment appears an equal number of times in each period. We now proceed by detailing each of our explored methods for estimating $\sigma_e^2$ and $\sigma_b^2$ based on the internal pilot data.

\subsection{Unblinded estimator}

The first of the methods we consider is an unblinded procedure. As noted, such an approach is typically less well favoured by regulatory agencies. However, though this may not always actually prove to be the case (see, e.g., \citet{friede2013}), one may anticipate its performance in terms of estimating the key variance components and provided desired operating characteristics to be preferable to that of the blinded procedures. This method therefore serves as a standard against which to assess the blinded estimators. Explicitly, this approach breaks the randomisation code and fits the linear mixed model~(\ref{eq2}) to the accrued data using REML estimation. With the REML estimates of $\sigma_e^2$ and $\sigma_b^2$ obtained, they are utilised in the re-estimation procedure as described above.

\subsection{Adjusted blinded estimator}

\citet{zucker1999} considered a blinded estimator for two-arm parallel trial designs based on an adjustment to the one-sample variance. \citet{golkowski2014} considered a similar unadjusted procedure for two-arm bioequivalence trials. Here, we consider a similar approach for multi-treatment crossover trials. Specifically, the following blinded estimators of the within and between person variances are used
	
	\begin{align*}
	\begin{split}
	\hat{\sigma}_{e}^2 &= \frac{1}{2(P - 1)(n_{\text{int}} - 1)} \sum_{j=2}^{P}\sum_{k=1}^{K} \sum_{i=1}^{n_{\text{int}}/K}(p_{ijk} - \bar{p}_j)^2 \\ & \qquad -\frac{n_{\text{int}}}{2K(P-1)(n_{\text{int}} - 1)}\sum_{j=2}^P\sum_{k=1}^{K}(\tau_{\text{d}(j,k)}^* - \tau_{\text{d}(j-1,k)}^*)^2,
	\end{split}\\
	\begin{split}
	\hat{\sigma}_{b}^2 &= \frac{1}{2}\left\{ \frac{1}{2(P - 1)(n_{\text{int}} - 1)} \sum_{j=2}^{P}\sum_{k=1}^{K} \sum_{i=1}^{n_{\text{int}}/K}(q_{ijk} - \bar{q}_j)^2 - \hat{\sigma}_{e}^2 \right.\\ & \qquad \qquad \left.- \frac{n_{\text{int}}}{2K(P-1)(n_{\text{int}} - 1)}\sum_{j=2}^P\sum_{k=1}^{K}(\tau_{\text{d}(j,k)}^* + \tau_{\text{d}(j-1,k)}^*)^2 \right. \\ & \qquad\qquad\qquad \left.+ \frac{2n_{\text{int}}}{D^2(n_{\text{int}}-1)}\left( \sum_{k=1}^K\tau_{\text{d}(1,k)} \right)^2\right\},
	\end{split}
	\end{align*}
	for specified $\tau_d^*$, $d=0,\dots,D-1$, with $\tau_0^*=0$, where
	
	\begin{align*}
	p_{ijk} &= y_{ijk} - y_{ij-1k},\\
	q_{ijk} &= y_{ijk} + y_{ij-1k},\\
	\bar{p}_{j} &= \frac{1}{n_{\text{int}}} \sum_{k=1}^{K}\sum_{i=1}^{n_{\text{int}}/K}p_{ijk},\\
	\bar{q}_{j} &= \frac{1}{n_{\text{int}}} \sum_{k=1}^{K}\sum_{i=1}^{n_{\text{int}}/K}q_{ijk}.
	\end{align*}
	
	In the Appendix, we show that if $\tau_{d}^*=\tau_d$ for $d=1,\dots,D-1$ then $\text{E}(\hat{\sigma}_{e}^2) = \sigma_e^2$ and $\text{E}(\hat{\sigma}_{b}^2) = \sigma_b^2$, and thus $\hat{\sigma}_{e}^2$ and $\hat{\sigma}_{b}^2$ are unbiased estimators for $\sigma_e^2$ and $\sigma_b^2$ respectively. This is the reason for our restrictions on the employed randomisation scheme (which assumes $n_1=\dots=n_K=n_{\text{int}}/K$ at the interim reassessment), and the employed sequences (which are assumed to be balanced for period). The above estimator could be used when there is imbalance in the number of patients allocated to each sequence, or without making this restriction on the sequences, but results on the expected values of the variance components would have a more complex form. It is therefore advantageous to ensure an equal number of patients are allocated to each sequence, and also logical to utilise period balanced sequences. We also view is as sensible therefore to explore the performance of the estimators in this case.

It is also important to assess the sensitivity of the performance of these estimators to the choice of the $\tau_{d}^*$, hoping for it to have negligible impact as in analogous procedures for other trial settings \citep{kieser2000}. Adapting previous works (see, e.g., \citet{kieser2003,zucker1999,gould1992}), we assess this procedure for $\tau_{d}^*=0$, and $\tau_{d}^*=\delta$, $d=1,\dots,D-1$, and refer to these henceforth as the null adjusted and alternative adjusted re-estimation procedures respectively. 

Note that one limitation of this approach in practice is that there is no guarantee that the above value for $\hat{\sigma}_{b}^2$ will be positive. Therefore, we actually re-evaluate the required sample size as $\text{N}\{\hat{\sigma}_{e}^2,\max(0,\hat{\sigma}_{b}^2)\}$. For the examples provided in the Appendix, we demonstrate that the above procedure still performs well despite this inconvenience. Moreover, in certain routinely faced scenarios, as will be discussed shortly, the value of $\sigma_{b}^2$ is inconsequential and this issue therefore no longer exists. However, in general this must be kept in mind when considering using this procedure for sample size re-estimation.

\subsection{Blinded estimator following block randomisation}

The above re-estimation procedures are explored within the context of a simple randomisation scheme that only ensures an equal number of patients are allocated to each sequence prior to the interim re-assessment. In contrast, the final blinded estimator we consider exploits the advantages block randomisation can bring, extending the methodology presented in \citet{xing2005} for parallel arm trials to crossover studies.

We suppose that patients are allocated to sequences in $B$ blocks, each of length $n_B$ (with these values chosen such that $Bn_B = n_{\text{int}}$). We recategorise our data as $y_{ijb}$, the response from patient $i=1,\dots,n_B$, in period $j$, in block $b$. Then, the following blinded estimators are used to recalculate the required sample size
	\begin{align*}
	\hat{\sigma}_{e}^2 &= \frac{1}{2(P - 1)(n_{\text{int}} - B)}\sum_{j=2}^{P}\sum_{b=1}^{B}\sum_{i=1}^{n_B}(p_{ijb} - \bar{p}_{jb})^2,\\
	\hat{\sigma}_{b}^2 &= \frac{1}{2}\left\{ \frac{1}{2(P - 1)(n_{\text{int}} - B)}\sum_{j=2}^{P}\sum_{b=1}^{B}\sum_{i=1}^{n_B}(q_{ijb} - \bar{q}_{jb})^2 - \hat{\sigma}_{e}^2 \right\},
	\end{align*}
	where
	\begin{align*}
	p_{ijb}&=y_{ijb}-y_{ij-1b},\\
	q_{ijb} &= y_{ijb} + y_{ij-1b},\\
	\bar{p}_{jb} &= \frac{1}{n_{B}} \sum_{i=1}^{n_B}p_{ijb},\\
	\bar{q}_{jb} &= \sum_{i=1}^{n_B}\sum_{i=1}^{n_{B}}q_{ijb}.
	\end{align*}
	
	In the Appendix, provided that an equal number of patients are allocated to each of a set of period balanced sequences, these are also shown to be unbiased estimators for $\sigma_{e}^2$ and $\sigma_{b}^2$. Note though that as above, we must actually re-estimate $N$ using $\text{N}\{\hat{\sigma}_{e}^2,\max(0,\hat{\sigma}_{b}^2)\}$. Additionally, when using block randomisation, the actual sample size used by a trial may differ from $\hat{N}$, if it is not divisible by the block length $n_B$.

\section{Simulation study}

	\subsection{Motivating examples}
	
	We present results for three motivating examples based on real crossover trials. Example 1 is described in Section 3.2, with Examples 2 and 3 discussed in the Appendix, where their associated results are also presented. Among the three examples we consider settings with a range of required sample sizes, utilising complete block, incomplete block, and extra-period designs. This allows us to provide a thorough depiction of the performance of the various estimators in a wide range of realistic trial design settings.

\subsection{Example 1: TOMADO}\label{sec:tomado}

First, we assess the performance of the various re-estimation procedures using the TOMADO trial as motivation. TOMADO compared the clinical effectiveness of a range of mandibular devices for the treatment of obstructive sleep-apnea hypopnea. Precise details can be found in \citet{quinnell2014}. Briefly, TOMADO was a four-treatment four-period crossover trial, with patients allocated treatment sequences using two Williams squares. The data for the outcome Epworth Sleepiness Scale was to be analysed using linear mixed model~(\ref{eq2}), with the following hypotheses tested
\[ H_{0d} : \tau_d \ge 0, \qquad H_{1d} : \tau_d < 0, \qquad d=1,\dots,D-1, \]
since a reduction in the Epworth Sleepiness Scale score is indicative of an efficacious treatment. Consequently, the null hypotheses were to be rejected if $T_d<-e$, using the value of $e$ determined as above.

Following the methodology described in the Appendix, we can demonstrate that when complete-block period-balanced sequences are used for treatment allocation, that the required sample size has no dependence upon the between person variance $\sigma_b^2$. Explicitly, we have
	\begin{equation}\nonumber
	\text{N}(\sigma_e^2,\sigma_b^2)\equiv\text{N}(\sigma_e^2) = \frac{2\sigma_e^2(z_{1-\alpha_*} + z_{1-\beta})^2}{\delta^2},
	\end{equation}
	where $\alpha_*$ is defined in the Appendix. See \citet{jones2014}, for an alternative derivation of this formula. This substantially simplifies the re-estimation procedure, as we only need to provide a value for $\sigma_e^2$, and do not require use of the estimators for $\sigma_b^2$.

TOMADOs complete case analysis estimated the following values for the various components of the linear mixed model (1)
\[ \hat{\mu}_0 = 10.65, \ \hat{\pi}_2 = -0.77, \ \hat{\pi}_3 = -0.96, \ \hat{\pi}_4 = -0.55,\]
\[ \hat{\tau}_1 = -1.51, \ \hat{\tau}_2 = -2.15, \ \hat{\tau}_3 = -2.37, \ \hat{\sigma}_e^2 = 6.51,\ \hat{\sigma}_b^2 = 10.12. \]
Therefore, for $\sigma_e^2=\hat{\sigma}_e^2$, the trials planned recruitment of 72 patients would have conferred power of 0.8 at a significance level of 0.05 for $\delta=-1.24$. Consequently, we set $\beta=0.2$ and $\alpha=0.05$ throughout. In the main manuscript, we additionally take $\delta=-1.24$ and $\sigma_b^2=10.12$ always. The effect of other underlying values for $\delta$ and $\sigma_b^2$ is considered in the Appendix. In contrast, whilst we focus here on the case with $\sigma_e^2=6.51$, we also consider the influence of alternative values for this parameter. When simulating data we take $\mu_0 = 10.65$, $\pi_2 = -0.77$, $\pi_3 = -0.96$, and $\pi_4 = -0.55$. However, the effect of other period effects is discussed in Section~\ref{disc} and in the Appendix.

We explore the performance of the procedures under the global null hypothesis ($\tau_1=\tau_2=\tau_3=0$), when only treatment one is effective ($\tau_1=\delta,\tau_2=\tau_3=0$), when treatments one and two are effective ($\tau_1=\tau_2=\delta,\tau_3=0$), under the global alternative hypothesis ($\tau_1=\tau_2=\tau_3=\delta$), and under what we refer to henceforth as the observed treatment effects ($\tau_1=-1.51, \tau_2=-2.15, \tau_3=-2.37$). For simplicity, we assume a single Latin square was used for treatment allocation, and set $n_{\text{max}}=1000$ so that there is no practical upper limit on the allowed sample size. In all cases, the average result for a particular design and analysis scenario was determined using 100,000 trial simulations.

\subsection{Distributions of $\hat{\sigma}_e^2$ and $\hat{N}$}

First, the performance of the re-estimation procedures was explored for the parameters listed in Section~\ref{sec:tomado}, with $\sigma_e^2=6.51$, and $n_{\text{int}}\in \{8, 16, 24, 32, 40\}$. The resulting distributions of $\hat{\sigma}_e^2$, the interim estimate of $\sigma_e^2$, are shown in Figure 1 via the median, lower and upper quartiles in each instance. Additionally, Figure 2 depicts the equivalent results for the distribution of $\hat{N}$, the interim re-estimated value for $N$. The results are grouped according to the timing of the re-estimation and by the true value of the treatment effects. Note that $n_B=4$ is only considered for values of $n_{\text{int}}$ which allows an equal number of patients to be allocated to each sequence by the interim analysis.

The median value of $\hat{\sigma}_e^2$ for the unblinded procedure is always close to, but typically slightly less than, the true value $\sigma_e^2$. The same statement holds for the block randomisation procedure with $n_B=2$ or 4. However, whilst this is true for the adjusted procedures under the global null hypothesis, it is not otherwise always the case. In particular, both perform poorly for the observed treatment effects.
	
	As would be anticipated, the alternative adjusted procedure has lower median values for $\hat{\sigma}_e^2$ than the null adjusted procedure. Moreover, using the block randomised re-estimation procedure with $n_B=4$ seems to improve performance over $n_B=2$, both in terms of the median value of $\hat{\sigma}_e^2$, and by imparting a smaller interquartile range for $\hat{\sigma}_e^2$.

The results for $\hat{N}$ mirror those for $\hat{\sigma}_e^2$. Thus $\hat{N}$ is larger for the adjusted estimators under the observed treatment effects, but otherwise the distributions are comparable.

Increasing the value of $n_{\text{int}}$ reduces the interquartile range for $\hat{\sigma}_e^2$ and $\hat{N}$ for each procedure, and results in median values closer to the truth, as would be expected. Finally, we observe that the interquartile range for the unblinded procedure is often smaller than that of its adjusted or block randomisation counterparts.

\subsection{Familywise error-rate and power}

For the scenarios from Section 3.3 that were not conducted under the observed treatment effects, the estimated FWER and power were also recorded. The results are displayed in Table 1.

The FWER for each of the procedures is usually close to the nominal level, with a maximal value of 0.052 for the unblinded procedure with $n_{\text{int}}=32$. The adjusted procedures arguably have the smallest inflation across the considered values of $n_{\text{int}}$.

In most cases the re-estimation procedures attain a power close to the desired level. Of the adjusted procedures, the null adjusted has a larger power, as would be anticipated given our observations on $\hat{\sigma}_e^2$ and $\hat{N}$ above. In fact, the null adjusted method conveys the highest power for each value of $n_{\text{int}}$. The power of the block randomised procedures is typically similar to that of the alternative adjusted method. In addition, whether only treatment one, treatments one and two, or all three treatments are effective has little effect on the power.

There is no clear to trend as to the effect of increasing $n_{\text{int}}$ on the FWER, however it leads in almost all instances to an improvement in power. Finally, increasing the value for $n_B$ in the block randomisation procedure increases power as would be predicted.

\subsection{Influence of $\sigma_e^2$}
	
	In this section, we consider the influence of the value of $\sigma_e^2$ on the performance of our re-estimation procedures. Specifically, whilst we know that increasing $\sigma_e^2$ will increase the required sample size, we would like to assess the effect this has upon the ability of the methods to control the FWER and attain the desired power.
	
	Figures 3 and 4 respectively present our results on the FWER and power of the various re-estimation procedures when $n_{\text{int}}\in\{16,32\}$ for several values of $\sigma_e^2\in[0.25(6.51),4(6.51)]$, under the global null and alternative hypotheses respectively.
	
	Arguably, we observe that the FWER is more variable for smaller values of $\sigma_e^2$, with it changing little for several of the procedures when $\sigma_e^2>10$. There is additionally some evidence to suggest that increasing the value of $n_{\text{int}}$ reduces the overall effect $\sigma_e^2$ has on the FWER.
	
	For the power, as would be anticipated, the re-estimation procedures are over-powered when $n_{\text{int}}=32$ and $\sigma_e^2$ is small. Moreover, increasing the value of $n_{\text{int}}$ universally increases the power. Finally, as $\sigma_e^2$ increases beyond approximately $\sigma_e^2=5$, for both considered values of $n_{\text{int}}$, there is little change in power.

\subsection{Sample size inflation factor}

Whilst the above suggests the overall performance of the re-estimation is good, there are several simple refinements that can be implemented to improve the observed results.

One such refinement, to help ensure the power provided by the re-estimation procedures is at least the desired $1-\beta$, is to utilise a sample size inflation factor as originally proposed by \citet{zucker1999}. With it, the value of $\hat{N}$ as determined using the arguments above, is enlarged by the following factor

\begin{equation}\nonumber
\left(\frac{t_{1-\alpha,\nu_{n_\text{int}}} + t_{1-\beta,\nu_{n_\text{int}}}}{z_{1-\alpha} + z_{1-\beta}}\right)^2.
\end{equation}

Of course, one must be careful that the new implied sample size does not exceed any specified value of $n_{\text{max}}$. However, this factor has then been shown to improve the performance of re-estimation procedures in both superiority \citep{zucker1999}, non-inferiority \citep{friede2013}, and two-treatment bioequivalence trials \citep{golkowski2014}.

Figure 5 displays its effect in the context of our multi-treatment crossover trials. Explicitly, the power of the various re-estimation procedures under the global alternative hypothesis, for $n_{\text{int}}\in\{8,16,24,32,40\}$ and $\sigma_e^2=6.51$, is shown with and without the use of the inflation factor. For the unblinded, null adjusted, and block randomised method with $n_B=4$, the inflation factor increases power to above the desired level in every instance. Consequently, this simple inflation factor appears once more to be an effective adjustment to the basic procedures.

\section{Discussion}\label{disc}

In this article, we have developed and explored several methods for the interim re-assessment of the sample size required by a multi-treatment crossover trial. Our methodology is applicable to any trial analysed using the linear mixed model~(\ref{eq2}), when there is equal participant allocation to a set of period balanced sequences. Thus whilst adapting the work of \citet{golkowski2014} would be advisable in the case of an AB/BA superiority trial, given that it does not require the use of simulation, our methods are pertinent to a broader set of crossover designs. Indeed, they are as readily applicable to multi-treatment superiority trials as they are ones for establishing bioequivalence.

We explored performance via three motivating examples, allowing consideration of settings with different types of sequences and a range of required sample sizes. Overall, the results presented here for the TOMADO trial are similar to those provided in the Appendix for Examples 2 and 3. However, larger inflation to the FWER was observed in Example 2, most likely as a consequence of its associated smaller sample sizes. Nonetheless, the methods were found to provide desirable power characteristics with negligible inflation to the FWER in many settings. In particular, the blinded procedures provided comparable operating characteristics to the unblinded procedure, and thus can be considered viable alternatives. Following results for parallel arm trials \citep{friede2013}, the null adjusted blinded estimator arguably performed better than the other estimators in that its typical over-estimation of the variance at interim led to the desired power being achieved more often. We may therefore tentatively suggest the null adjusted blinded estimator to be the preferred approach in this setting.

Our findings indicate that for each of the re-estimation procedures, the choice of $\delta$ and the underlying values of $\sigma_e^2$ and $\sigma_b^2$ often have little effect upon the FWER and power. We may be reassured therefore that the performance of the procedures should often be relatively insensitive to the design parameters. On a similar note, it is important to recognise that one cannot be certain when utilising these methods that the value of the period effects will not influence the performance of the re-estimation procedures. Whilst the final analysis should be asymptotically invariant to period effects, in finite samples it may influence the results of the hypothesis tests. Intuitively though one would not anticipate this effect to be large, nor would one routinely expect large period effects in many settings. In the Appendix, simulations to explore this are presented for the TOMADO example. The results indicate that there is little evidence to suggest the value of the period effects influences the performance of the re-estimation procedures. Trialists must be mindful however that this cannot be guaranteed, and should therefore be investigated.

We also considered the utility of a simple sample size inflation factor in ensuring the power reaches the desired level. Ultimately, we demonstrated that this was an effective extension to the basic re-estimation procedures. Though the observed inflation to the FWER of our procedures was often small, if more strict control is desired, a crude $\alpha$-level adjustment procedure can also be utilised. For a particular re-estimation scenario, the values of $\sigma_e^2$ and $\sigma_b^2$, $\sigma_{e,\text{max}}^2$ and $\sigma_{b,\text{max}}^2$ say, which maximise the inflation to the FWER under the global null hypothesis can be determined via a two dimensional search. Then, the significance level used in the analysis of the trial can be adjusted to the $\alpha_{\text{adj}}$ that confers a FWER of $\alpha$ for this $\sigma_{e,\text{max}}^2$, $\sigma_{b,\text{max}}^2$ pair, according to further simulations. This may be useful in practice if the inflation is large for a particular trial design scenario of interest.

It is important to note the seemingly inherent advantages and disadvantages of the various re-estimation procedures. The adjusted estimator is perhaps the most constrained of those considered; requiring an equal number of patients to be allocated to each sequence for any non-zero adjustment to be reasonable. This is particularly troubling because of the possibility of patient drop-out. 

The estimator following block randomisation does not necessitate equal allocation to sequences (though its performance was considered here only when this was the case), but could also fall foul of patient drop-out that would prevent the estimation of the within person variance for each block. It also requires block randomisation, and could not be used with a more simple randomisation scheme if this was desired. The unblinded estimator of course suffers from none of these problems, but as discussed may be looked upon less favourably by regulators.

Finally, note that in conducting our work we also considered the performance of two re-estimation procedures based on methodology for the clustering of longitudinal data \citep{fraley2003,genolini2015}. The motivation for this came from the Expectation-Maximisation algorithm approaches of \citet{gould1992} for parallel two-arm, and \citet{kieser2000} for parallel multi-arm, studies. These methods may seem appealing, as they are blinded, under certain assumptions can produce unbiased estimates of the variance parameters, do not require specification of any adjustment, and in theory should be able to more readily handle small amounts of missing data. However, we found that they routinely vastly under-estimated the size of within person variance, resulting in substantially lower power than that attained by the other re-estimation procedures. Accordingly, especially given the associated concerns about the appropriateness of an Expectation-Maximisation algorithm for blinded sample size re-estimation \citep{friede2002}, we would not recommend re-estimation be performed based on a clustering based approach.

In conclusion, following findings for other trial design settings, blinded estimators can be used for sample size re-estimation in multi-treatment crossover trials. The operating characteristics of any chosen procedure should of course be assessed pre-trial through a comprehensive simulation study. But, often, investigators can hope to find that the likelihood of correctly powering their study when there is pre-trial uncertainty over the within and between person variances can be enhanced.

\section*{Acknowledgements}
The authors would like to thank the two anonymous reviewers whose comments helped to substantially improve the quality of this article. This work was supported by the Medical Research Council [grant number MC\_UU\_00002/3 to M.J.G. and A.P.M., and grant number MC\_UU\_00002/6 to J.M.S.W.].
\vspace*{1pc}

\begin{figure}[htb]
	\begin{center}\label{sigmae2hat}
		\includegraphics[width = 15cm]{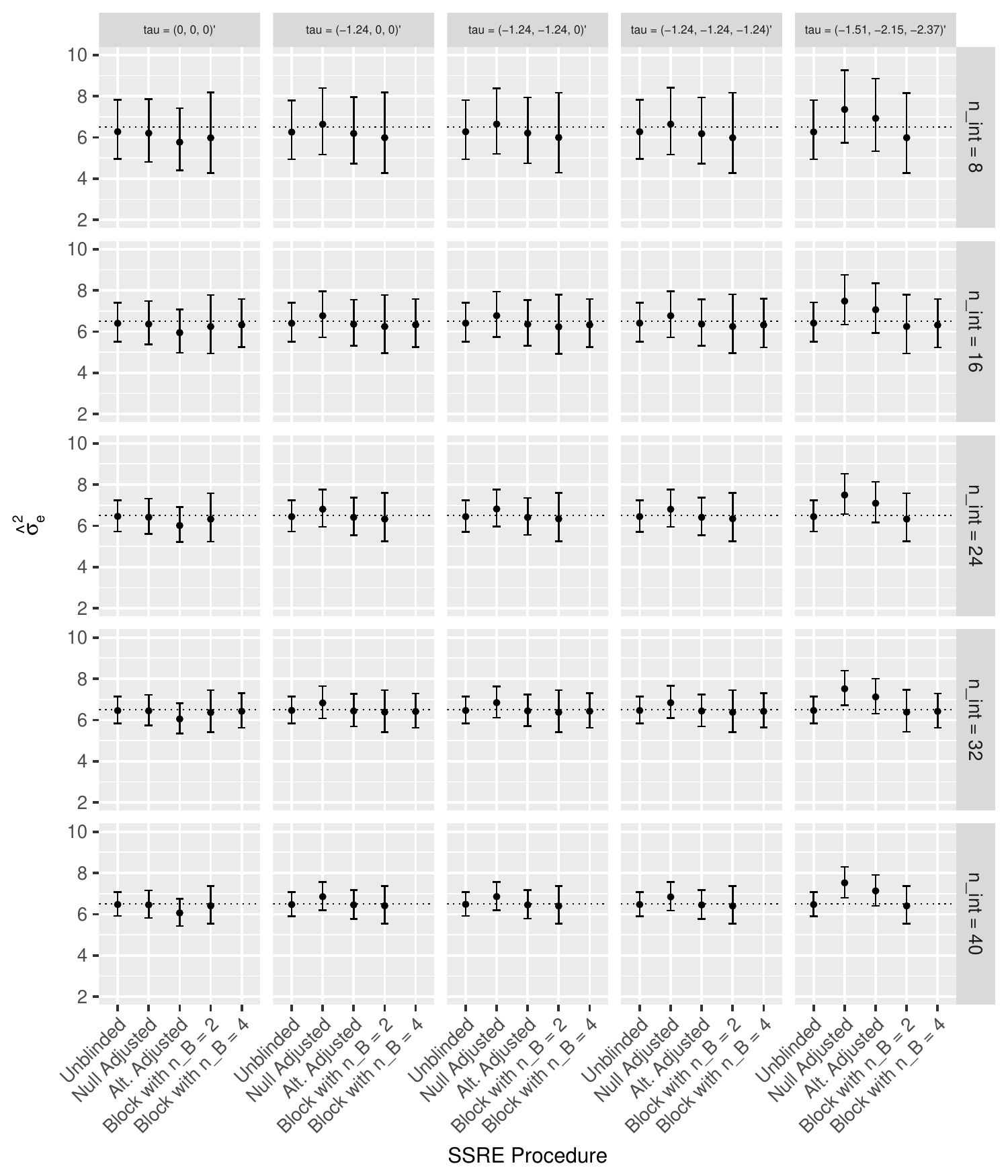}
		\caption{The distribution of $\hat{\sigma}_e^2$ is shown for each of the re-estimation procedures for several values of $\boldsymbol{\tau}$, and several values of $n_{\text{int}}$, for Example 1. Precisely, for each scenario, the median, lower and upper quartile values of $\hat{\sigma}_e^2$ across the simulations are given. The dashed line indicates the true value of $\sigma_e^2$.}
	\end{center}
\end{figure}

\begin{figure}[htb]
	\begin{center}\label{Nhat}
		\includegraphics[width = 15cm]{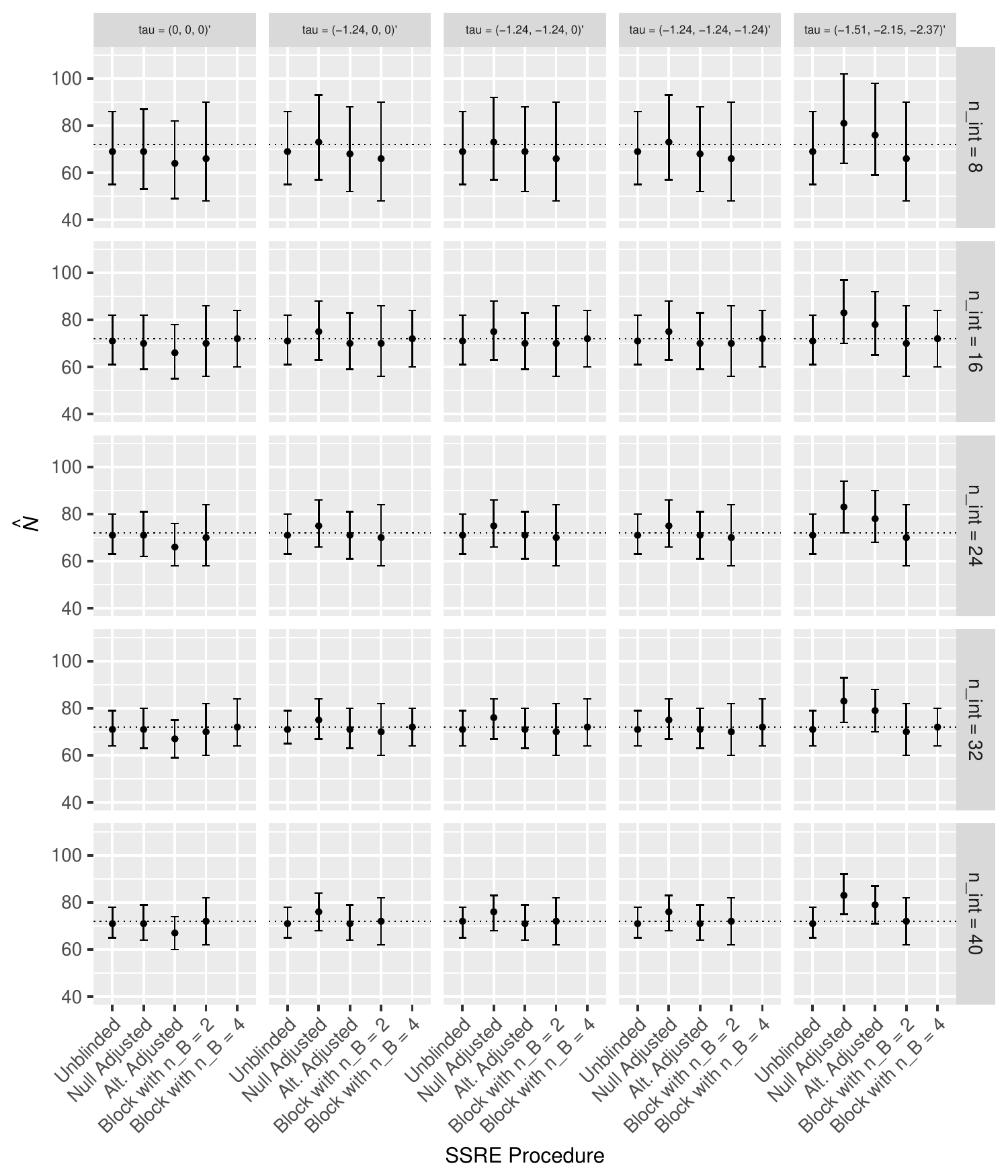}
		\caption{The distribution of $\hat{N}$ is shown for each of the re-estimation procedures for several values of $\boldsymbol{\tau}$, and several values of $n_{\text{int}}$, for Example 1. Precisely, for each scenario, the median, lower and upper quartile values of $\hat{N}$ across the simulations are given. The dashed line indicates the true required value of $N$.}
	\end{center}
\end{figure}

\begin{figure}[htb]
	\begin{center}\label{sigma_e_FWER}
		\includegraphics[width = 15cm]{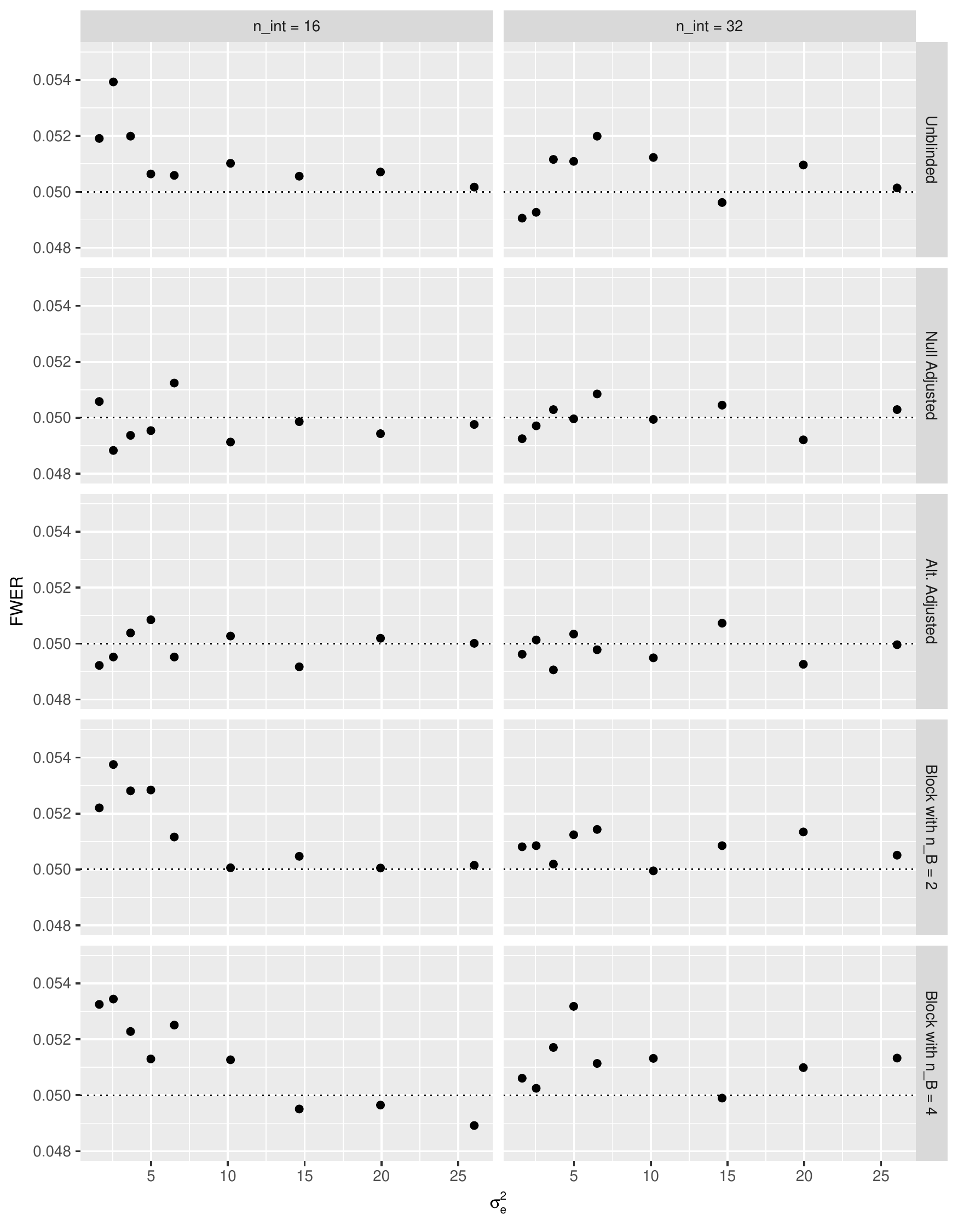}
		\caption{The simulated familywise error-rate (FWER) is shown under the global null hypothesis for each of the re-estimation procedures when $n_{\text{int}}\in\{16,32\}$, as a function of the within person variance $\sigma_e^2$, for Example 1. The Monte Carlo error is approximately 0.0007 in each instance. The dashed line indicates the desired value of the FWER.}		
	\end{center}
\end{figure}

\begin{figure}[htb]
	\begin{center}\label{sigma_e_power}
		\includegraphics[width = 15cm]{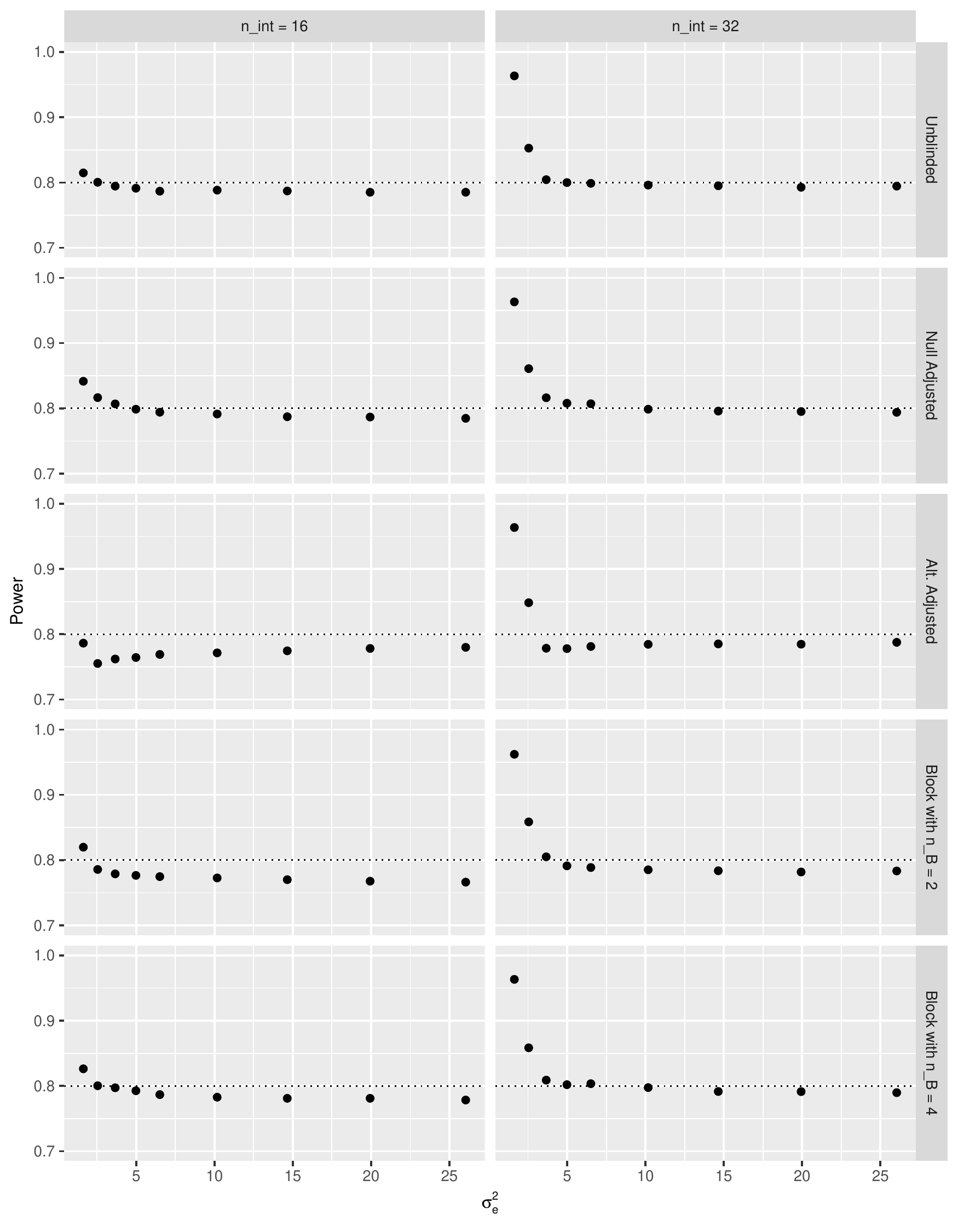}
		\caption{The simulated power is shown under the global alternative hypothesis for each of the re-estimation procedures when $n_{\text{int}}\in\{16,32\}$, as a function of the within person variance $\sigma_e^2$, for Example 1. The Monte Carlo error is approximately 0.0013 in each instance. The dashed line indicates the desired value of the power.}
	\end{center}
\end{figure}

\begin{figure}[htb]
	\begin{center}\label{inflationfactor}
		\includegraphics{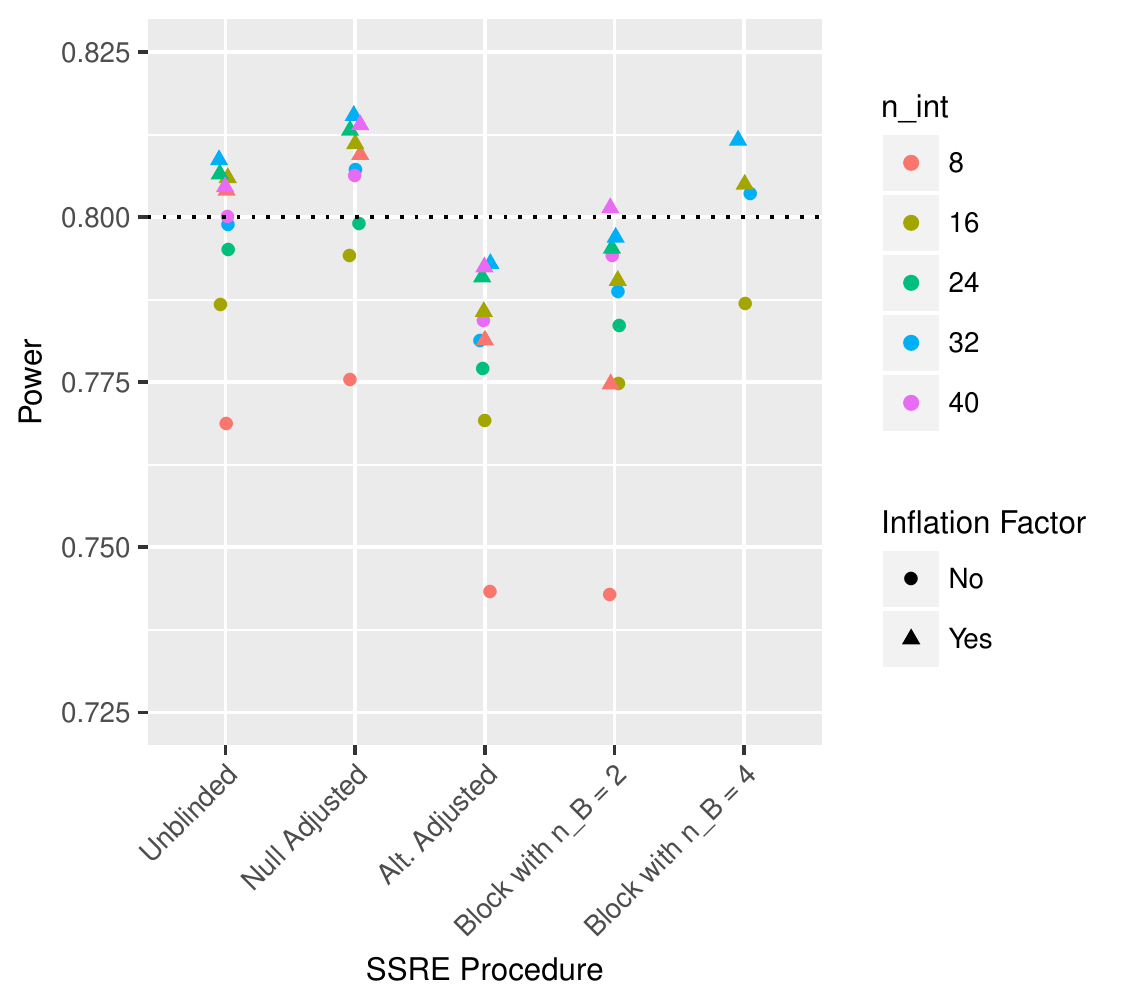}
		\caption{The influence of the considered inflation factor upon the power of the re-estimation procedures under the global alternative hypothesis is shown for several values of $n_{\text{int}}$, for Example 1. The dashed line indicates the desired value of the power.}
	\end{center}
\end{figure}

\appendix

\section{Deriving $\text{N}(\sigma_e^2,\sigma_b^2)$}
	
	In this section, we elaborate on how a formulae for the sample size required by a trial, $\text{N}(\sigma_e^2,\sigma_b^2)$, can be specified when $\sigma_e^2$ and $\sigma_b^2$ are known, a set of $K$ sequences have been chosen, and $n_1=\dots=n_K$. First, we focus on the set up from Section 2.1, before briefly describing adjustments for other testing scenarios.
	
	We begin by denoting the linear mixed model for our vector of observations $\boldsymbol{y}$ by $\boldsymbol{y}=\boldsymbol{X}\boldsymbol{\beta}+\boldsymbol{Z}\boldsymbol{b}+\boldsymbol{\epsilon}$. Here, it is important to note that the particular form of the matrices $\boldsymbol{X}$ and $\boldsymbol{Z}$ is dependent on the sample size and the chosen sequences. Then, the generalised least squares estimate for $\boldsymbol{\beta}$, $\hat{\boldsymbol{\beta}}$, is given by
	\[ \hat{\boldsymbol{\beta}} = (\boldsymbol{X}^\top\boldsymbol{\Sigma}^{-1}\boldsymbol{X})^{-1}\boldsymbol{X}^\top\boldsymbol{\Sigma}^{-1}\boldsymbol{y}, \]
	where $\boldsymbol{\Sigma}=\boldsymbol{Z}\boldsymbol{G}\boldsymbol{Z}^\top+\boldsymbol{R}$ is known. Precisely, by our choice of covariance structure implied by linear mixed model (1), $\boldsymbol{\Sigma}$ will be an $NP\times NP$ block diagonal matrix, consisting of $P\times P$ blocks given by $\sigma_e^2\boldsymbol{I}_P + \sigma_b^2\boldsymbol{J}_P$. In addition, $\boldsymbol{\text{Var}}(\hat{\boldsymbol{\beta}})=(\boldsymbol{X}^\top\boldsymbol{\Sigma}^{-1}\boldsymbol{X})^{-1}$, will also be known. Finally, $\hat{\boldsymbol{\beta}}$ is an unbiased estimate of $\boldsymbol{\beta}$. Thus the vector of test statistics $\boldsymbol{Q}=(Q_1,\dots,Q_{D-1})^\top$, where
	\[ Q_d = \frac{\hat{\tau}_d}{\{\text{Var}(\hat{\tau}_d)\}^{1/2}},\qquad d=1,\dots,D-1, \]
	has the following multivariate normal distribution
	\[ \boldsymbol{Q} \sim N\left\{\boldsymbol{\tau}\circ\boldsymbol{I}^{1/2}, \boldsymbol{Diag}(\boldsymbol{I}^{1/2})\boldsymbol{\text{Var}}(\hat{\boldsymbol{\tau}})\boldsymbol{Diag}(\boldsymbol{I}^{1/2})\right\}. \]
	Here, $\boldsymbol{\tau}=(\tau_1,\dots,\tau_{D-1})^\top$, and $\boldsymbol{I}=(I_{1},\dots,I_{D-1})^\top$ with $I_{d}=\{\text{Var}(\hat{\tau}_d)\}^{-1}$. Furthermore, $\boldsymbol{Diag}(\boldsymbol{v})$ is the matrix formed by placing the elements of the vector $\boldsymbol{v}$ along the leading diagonal, and we take the convention that $\{(v_1,\dots,v_m)^\top\}^{1/2}=(v_1^{1/2},\dots,v_m^{1/2})^\top$.
	
	Using the distribution of $\boldsymbol{Q}$, we can control the FWER to a desired level $\alpha$ for our hypotheses of interest by rejecting $H_{0d}$ if $Q_d>e$, for the $e$ which is the solution of
	\begin{equation*}
	1 - \alpha = \Phi_{D-1}\{(e,\dots,e)^\top,\text{Var}(\boldsymbol{Q})\},
	\end{equation*}
	where $\Phi_{M}\{\textbf{x},\Lambda\}$ is the $M$-dimensional cumulative distribution function of a central multivariate normal distribution with covariance matrix $\Lambda$.
	The power to reject $H_{01}$ when $\tau_1=\delta$ is then
	\[ \Phi_1\{\delta I_1^{1/2}-\Phi^{-1}_1(1-\alpha_*,1),1\}, \]
	for $\alpha_*=1-\Phi_1(e,1)$. Consequently, to have power of $1-\beta$ we must have
	\[ \Phi_1\{\delta I_1^{1/2}-\Phi^{-1}_1(1-\alpha_*,1),1\} = 1-\beta. \]
	By deriving the explicit form of $I_1$ we can determine its dependence upon the sample size $N$, which allows the above to be arranged and our final formula $\text{N}(\sigma_e^2,\sigma_b^2)$ specified.
	
	For example, as discussed in Section 3.2, in the case where complete block sequences that are balanced for period are utilised, it is well known that $I_1=N/(2\sigma_e^2)$ (Jones and Kenward, 2014). Thus, we can rearrange the above to acquire our previously specified formula
	\begin{equation}\nonumber
	\text{N}(\sigma_e^2,\sigma_b^2)\equiv\text{N}(\sigma_e^2)= \frac{2\sigma_e^2\{\Phi^{-1}_1(1-\alpha_*,1) + \Phi^{-1}_1(1-\beta,1)\}^2}{\delta^2}\equiv\frac{2\sigma_e^2(z_{1-\alpha_*} + z_{1-\beta})^2}{\delta^2}.
	\end{equation}
	In the case where alternative hypotheses are to be assessed (e.g., a global test that foregoes many-to-one comparisons), provided testing is performed using effects from $\hat{\boldsymbol{\beta}}$, the above can easily be adapted to designate an appropriate sample size formula. In each instance, one specifies a vector of test statistics, the distribution of which can be derived using that of $\hat{\boldsymbol{\beta}}$. This allows an appropriate formulae for controlling the FWER to be provided. For example, in the case where
	\[ H_{0d} : \tau_d = 0,\qquad H_{1d} : \tau_d\neq0,\qquad d=1,\dots,D-1, \]
	we still utilise $\boldsymbol{Q}$ as defined above, but now reject $H_{0d}$ if $|Q_d|>e$, where $e$ is the solution of
	\begin{equation*}
	1 - \alpha/2 = \Phi_{D-1}\{(e,\dots,e)^\top,\text{Var}(\boldsymbol{Q})\}.
	\end{equation*}
	Then, a formulae for the power of interest can similarly be designated, which allows $N$ to either be specified explicitly, or permits an iterative search to be performed to determine its required value.
	
	For example, for the design scenario of Section 2.1, if we instead desire a familywise power of at least $1-\beta$ when $\boldsymbol{\tau}=(\delta,\dots,\delta)^\top$ (that is, power to reject at least one of $H_{01},\dots,H_{0D-1}$), our formula for the power becomes
	\[ 1 - \Phi_{D-1}\{(e-\delta I_1^{1/2},\dots,e-\delta I_{D-1}^{1/2})^\top,\text{Var}(\boldsymbol{Q})\}. \]
	Having derived $\boldsymbol{I}$ and $\text{Var}(\boldsymbol{Q})$ for any $N$, a one-dimensional search can then be performed to acquire the minimal $N$ such that the above is at least $1-\beta$.

\section{Adjusted blinded estimator}

In this section, we find the expected value of the adjusted blinded estimators of the within and between person variances discussed in the main paper.

First, note that for any $K$ sequences which are balanced for period
	\begin{equation}\label{perbal}
	\sum_{k=1}^{K} (\tau_{\text{d}(j,k)} - \tau_{\text{d}(j-1,k)}) = 0,
	\end{equation}
	for $j=2,\dots,P$, and
	\begin{equation}
	\sum_{k=1}^{K} \tau_{\text{d}(1,k)} = \dots = \sum_{k=1}^{K} \tau_{\text{d}(P,k)}.
	\end{equation}
Now note that for the linear mixed model (1) that
\begin{equation*}
\text{E}(p_{ijk}) = \pi_j - \pi_{j-1} + \tau_{\text{d}(j,k)} - \tau_{\text{d}(j-1,k)}.
\end{equation*}
In addition, observe that $\text{Cov}(p_{i_1j_1k_1},p_{i_2j_2k_2})=0$ unless $i_1=i_2$ and $k_1=k_2$. Furthermore
\begin{align*}
\text{Var}(p_{ijk}) &= \text{Cov}(y_{ijk}-y_{ij-1k},y_{ijk}-y_{ij-1k}),\\
&= \text{Cov}(y_{ijk},y_{ijk})-\text{Cov}(y_{ijk},y_{ij-1k})\\
& \qquad -\text{Cov}(y_{ij-1k},y_{ijk})+\text{Cov}(y_{ij-1k},y_{ij-1k}),\\
&= (\sigma_e^2+\sigma_b^2) - \sigma_b^2 - \sigma_b^2 + (\sigma_e^2+\sigma_b^2),\\
&= 2\sigma_e^2.
\end{align*}
Consequently
\begin{align*}
\text{E}\left( \bar{p}_j \right) &= \text{E}\left( \frac{1}{n_{\text{int}}} \sum_{k=1}^{K}\sum_{i=1}^{n_{\text{int}}/K} p_{ijk} \right),\\
&= \frac{1}{n_{\text{int}}} \sum_{k=1}^{K}\sum_{i=1}^{n_{\text{int}}/K} \text{E}(p_{ijk}),\\
&= \frac{1}{n_{\text{int}}} \sum_{k=1}^{K}\sum_{i=1}^{n_{\text{int}}/K} (\pi_j - \pi_{j-1} + \tau_{\text{d}(j,k)} - \tau_{\text{d}(j-1,k)}),\\
&= \frac{1}{n_{\text{int}}} n_{\text{int}}(\pi_j - \pi_{j-1}),\\
&= \pi_j - \pi_{j-1},
\end{align*}
where we have used Equation (3). Additionally
\begin{align*}
\text{Var}\left( \bar{p}_j \right) &= \text{Var}\left( \frac{1}{n_{\text{int}}} \sum_{k=1}^{K}\sum_{i=1}^{n_{\text{int}}/K} p_{ijk} \right),\\
&= \frac{1}{n_{\text{int}}^2} \text{Var}\left( \sum_{k=1}^{K}\sum_{i=1}^{n_{\text{int}}/K} p_{ijk} \right),\\
&= \frac{1}{n_{\text{int}}^2} \left\{ \sum_{k=1}^{K}\sum_{i=1}^{n_{\text{int}}/K} \text{Var}\left(p_{ijk}\right) + \sum_{k_1\neq k_2}\sum_{i_1 \neq i_2} \text{Cov}\left(p_{i_1jk_1},p_{i_2jk_2}\right) \right\},\\
&= \frac{1}{n_{\text{int}}^2} \left\{ \sum_{k=1}^{K}\sum_{i=1}^{n_{\text{int}}/K} \text{Var}\left(p_{ijk}\right) \right\},\\
&= \frac{1}{n_{\text{int}}^2} (2\sigma_e^2n_{\text{int}}),\\
&= \frac{2\sigma_e^2}{n_{\text{int}}}.
\end{align*}
Next, consider the following, which we call $\sigma^2_{\text{within}}$
\begin{align*}
\sigma^2_{\text{within}} &= \frac{1}{2(P - 1)(n_{\text{int}} - 1)} \sum_{j=2}^{P}\sum_{k=1}^{K} \sum_{i=1}^{n_{\text{int}}/K}(p_{ijk} - \bar{p}_j)^2,\\ 
&= \frac{1}{2(P - 1)(n_{\text{int}} - 1)} \sum_{j=2}^{P}\left(\sum_{k=1}^{K} \sum_{i=1}^{n_{\text{int}}/K}p_{ijk}^2 - n_{\text{int}}\bar{p}_j^2\right).
\end{align*}
Taking expectations we have
\begin{align*}
2(P - 1)(n_{\text{int}} - 1)\text{E}(\sigma^2_{\text{within}}) &= \sum_{j=2}^{P}\left\{\sum_{k=1}^{K} \sum_{i=1}^{n_{\text{int}}/K}\text{E}\left(p_{ijk}^2\right) - n_{\text{int}}\text{E}\left(\bar{p}_j^2\right)\right\},\\
&= \sum_{j=2}^{P}\left[\sum_{k=1}^{K} \sum_{i=1}^{n_{\text{int}}/K}\left\{ \text{Var}(p_{ijk}) + \text{E}(p_{ijk})^2 \right\} \right.\\ & \qquad \qquad \left. \phantom{\sum_{j=2}^{P}} -n_{\text{int}}\left\{\text{Var}(\bar{p}_j) + \text{E}(\bar{p}_j)^2 \right\} \right],\\
&= \sum_{j=2}^{P}\left[\sum_{k=1}^{K} \sum_{i=1}^{n_{\text{int}}/K}\left\{ 2\sigma_e^2 + (\pi_j - \pi_{j-1} + \tau_{\text{d}{(j,k)}} - \tau_{\text{d}{(j-1,k)}})^2\right\} \right.\\ & \qquad \qquad \left. - n_{\text{int}}\left\{\frac{2\sigma_e^2}{n_{\text{int}}} + (\pi_j - \pi_{j-1})^2 \right\} \right],\\
&= \sum_{j=2}^{P}\left\{ 2(n_{\text{int}} - 1)\sigma_e^2 + \sum_{k=1}^{K} \sum_{i=1}^{n_{\text{int}}/K}(\pi_j - \pi_{j-1} + \tau_{\text{d}{(j,k)}} - \tau_{\text{d}{(j-1,k)}})^2 \right.\\ & \qquad \qquad \left. \phantom{\sum_{j=2}^{P}} - n_{\text{int}}(\pi_j - \pi_{j-1})^2 \right\},\\
&= 2(P - 1)(n_{\text{int}} - 1)\sigma_e^2\\
& \qquad + \sum_{j=2}^{P}\left\{ \sum_{k=1}^{K} \sum_{i=1}^{n_{\text{int}}/K}(\pi_j - \pi_{j-1})^2 + \sum_{k=1}^{K} \sum_{i=1}^{n_{\text{int}}/K} (\tau_{\text{d}{(j,k)}} - \tau_{\text{d}{(j-1,k)}})^2 \right.\\ & \qquad \qquad \left. \phantom{\sum_{j=2}^{P}} + 2(\pi_j - \pi_{j-1})\sum_{k=1}^{K} \sum_{i=1}^{n_{\text{int}}/K} (\tau_{\text{d}{(j,k)}} - \tau_{\text{d}{(j-1,k)}}) \right.\\
& \qquad \qquad \left. \phantom{\sum_{j=2}^{P}} -  n_{\text{int}}(\pi_j - \pi_{j-1})^2 \right\},\\
&= 2(P - 1)(n_{\text{int}} - 1)\sigma_e^2\\
& \qquad + \sum_{j=2}^{P}\left\{ n_{\text{int}}(\pi_j - \pi_{j-1})^2 + \sum_{k=1}^{K} \sum_{i=1}^{n_{\text{int}}/K} (\tau_{\text{d}{(j,k)}} - \tau_{\text{d}{(j-1,k)}})^2 \right.\\
& \qquad \qquad \left. \phantom{\sum_{j=2}^{P}} -  n_{\text{int}}(\pi_j - \pi_{j-1})^2 \right\},\\
&= 2(P - 1)(n_{\text{int}} - 1)\sigma_e^2 + \sum_{j=2}^{P}\sum_{k=1}^{K} \sum_{i=1}^{n_{\text{int}}/K} (\tau_{\text{d}{(j,k)}} - \tau_{\text{d}{(j-1,k)}})^2,\\
&= 2(P - 1)(n_{\text{int}} - 1)\sigma_e^2 + \frac{n_{\text{int}}}{K}\sum_{j=2}^{P}\sum_{k=1}^{K} (\tau_{\text{d}{(j,k)}} - \tau_{\text{d}{(j-1,k)}})^2.
\end{align*}
Thus we have that
$$ \text{E}(\sigma^2_{\text{within}}) = \sigma_e^2 + \frac{n_{\text{int}}}{2K(P-1)(n_{\text{int}}-1)}\sum_{j=2}^{P}\sum_{k=1}^{K}(\tau_{\text{d}{(j,k)}} - \tau_{\text{d}{(j-1,k)}})^2, $$
and for our adjusted blinded estimator
\begin{equation}\nonumber
\hat{\sigma}_{e}^2 = \sigma^2_{\text{within}} - \frac{n_{\text{int}}}{2K(P-1)(n_{\text{int}}-1)}\sum_{j=2}^{P}\sum_{k=1}^{K}(\tau_{\text{d}{(j,k)}}^* - \tau_{\text{d}{(j-1,k)}}^*)^2,
\end{equation}
it is clear that if $\tau_{d}^*=\tau_d$ for $d=1,\dots,D-1$ then $\text{E}(\hat{\sigma}_{e}^2) = \sigma_e^2$, and $\hat{\sigma}_{e}^2$ is an unbiased estimator for $\sigma_e^2$ as claimed.

For the case where the sequences utilised for treatment allocation are additionally complete-block, the above estimator for $\sigma_e^2$ is all that is required by the adjusted blinded re-estimation procedures. When this is not the case however, we further require a blinded estimate of the between person variance $\sigma_b^2$. For this, define 
	\begin{align*}
	q_{ijk}&=y_{ij-1k}+y_{ijk},\\
	\bar{q}_{j}&=\frac{1}{n_{\text{int}}}\sum_{k=1}^K\sum_{i=1}^{n_{\text{int}}/K}q_{ijk}.
	\end{align*}
	Then, by a direct adaptation of the arguments above, and by using Equation (4), we can show that
	\begin{align*}
	\text{E}(q_{ijk}) &= \pi_{j-1}+\pi_j + \tau_{\text{d}(j-1,k)} + \tau_{\text{d}(j,k)},\\
	\text{Var}(q_{ijk}) &= 2(\sigma_e^2+2\sigma_b^2),\\
	\text{E}(\bar{q}_{j}) &= \pi_{j-1}+\pi_j + \frac{2}{D}\sum_{k=1}^K\tau_{\text{d}(1,k)},\\
	\text{Var}(\bar{q}_{j}) &= \frac{2(\sigma_e^2+2\sigma_b^2)}{n_{\text{int}}}.
	\end{align*}
	Now define
	\begin{equation*}
	\sigma_{\text{between}}^2=\frac{1}{2(P-1)(n_{\text{int}}-1)}\sum_{j=2}^P\sum_{k=1}^K\sum_{i=1}^{n_{\text{int}}/K}(q_{ijk}-\bar{q}_j)^2.
	\end{equation*}
	Modifying the derivations for $\sigma^2_{\text{within}}$, we have that
	\begin{align*}
	\text{E}(\sigma_{\text{between}}^2)&=\sigma_e^2 + 2\sigma_b^2 + \frac{n_{\text{int}}}{2K(P-1)(n_{\text{int}}-1)}\sum_{j=2}^P\sum_{k=1}^K(\tau_{\text{d}(j-1,k)} + \tau_{\text{d}(j,k)})^2\\ & \qquad +\frac{2n_{\text{int}}}{D^2(n_{\text{int}}-1)}\left( \sum_{k=1}^K\tau_{\text{d}(1,k)} \right)^2.
	\end{align*}
	We can thus then take our blinded estimate for $\sigma_b^2$, $\hat{\sigma}_b^2$, as
	\begin{align*}
	\hat{\sigma}_b^2 &= \frac{1}{2}\left\{ \sigma_{\text{between}}^2 - \hat{\sigma}_e^2 - \frac{n_{\text{int}}}{2K(P-1)(n_{\text{int}}-1)}\sum_{j=2}^P\sum_{k=1}^K(\tau_{\text{d}(j-1,k)}^* + \tau_{\text{d}(j,k)}^*)^2 \right. \\
	& \qquad \qquad \left. - \frac{2n_{\text{int}}}{D^2(n_{\text{int}}-1)}\left( \sum_{k=1}^K\tau_{\text{d}(1,k)}^* \right)^2 \right\}.
	\end{align*}
	Again, in the case that $\tau_{d}^*=\tau_d$ for $d=1,\dots,D-1$, this is an unbiased estimator for $\sigma_b^2$.

\section{Blinded estimator following block randomisation}

Here, we first demonstrate the forwarded blinded estimator for $\sigma_e^2$ following block randomisation is also unbiased. Throughout, the index $k$ above is essentially replaced by the index $b$. To begin, observe that
\begin{equation*}
\text{E}(p_{ijb}) = \pi_j - \pi_{j-1} + \tau_{\text{d}_B(j,b)} - \tau_{\text{d}_B(j-1,b)},
\end{equation*}
where $\text{d}_B(j,b)$ is the index of the treatment given to a patient in block $b$ in period $j$. As above, $\text{Cov}(p_{i_1j_1b_1},p_{i_2j_2b_2})=0$ unless $i_1=i_2$ and $b_1=b_2$, and $\text{Cov}(p_{ijb},p_{ijb}) = 2\sigma_e^2.$ Consequently

\begin{align*}
\text{E}\left( \bar{p}_{jb} \right) &= \text{E}\left( \frac{1}{n_B} \sum_{i=1}^{n_B} p_{ijb} \right),\\
&= \frac{1}{n_B} \sum_{i=1}^{n_B} \text{E}(p_{ijb}),\\
&= \frac{1}{n_B} \sum_{i=1}^{n_B} (\pi_j - \pi_{j-1} + \tau_{\text{d}_B(j,b)} - \tau_{\text{d}_B(j-1,b)}),\\
&= \frac{1}{n_B} n_B(\pi_j - \pi_{j-1} + \tau_{\text{d}_B(j,b)} - \tau_{\text{d}_B(j-1,b)}),\\
&= \pi_j - \pi_{j-1}+ \tau_{\text{d}_B(j,b)} - \tau_{\text{d}_B(j-1,b)},\\
\text{Var}\left( \bar{p}_{jb} \right) &= \text{Var}\left( \frac{1}{n_B} \sum_{i=1}^{n_B} p_{ijb} \right),\\
&= \frac{1}{n_B^2} \text{Var}\left( \sum_{i=1}^{n_B} p_{ijb} \right),\\
&= \frac{1}{n_B^2} \left\{ \sum_{i=1}^{n_B} \text{Var}\left(p_{ijb}\right) + \sum_{i_1 \neq i_2} \text{Cov}\left(p_{i_1jb},p_{i_2jb}\right) \right\},\\
&= \frac{1}{n_B^2} \left\{ \sum_{i=1}^{n_B} \text{Var}\left(p_{ijb}\right) \right\},\\
&= \frac{1}{n_B^2} (2\sigma_e^2n_B),\\
&= \frac{2\sigma_e^2}{n_B}.
\end{align*}

Next consider the proposed blinded estimator

\begin{align*}
\hat{\sigma}_e^2 &= \frac{1}{2(P - 1)(n_{\text{int}} - B)} \sum_{j=2}^{P}\sum_{b=1}^{B} \sum_{i=1}^{n_B}(p_{ijb} - \bar{p}_{jb})^2,\\
&= \frac{1}{2(P - 1)(n_{\text{int}} - B)} \sum_{j=2}^{P}\left[\sum_{b=1}^{B} \left(\sum_{i=1}^{n_B}p_{ijb}^2 - n_B\bar{p}_{jb}^2\right)\right].
\end{align*}

We have

\begin{align*}
2(P - 1)(n_{\text{int}} - B)\text{E}(\hat{\sigma}_e^2) &= \sum_{j=2}^{P}\left\{\sum_{b=1}^{B} \sum_{i=1}^{n_B}\text{E}\left(p_{ijb}^2\right) - n_B\sum_{b=1}^{B}\text{E}\left(\bar{p}_{jb}^2\right)\right\},\\
&= \sum_{j=2}^{P}\left[\sum_{b=1}^{B} \sum_{i=1}^{n_B}\left\{ \text{Var}(p_{ijb}) + \text{E}(p_{ijb})^2 \right\} \right.\\ & \qquad \qquad \left. \phantom{\sum_{j=2}^{P}} -n_{B}\sum_{b=1}^{B}\left\{\text{Var}(\bar{p}_{jb}) + \text{E}(\bar{p}_{jb})^2 \right\} \right],\\
&= \sum_{j=2}^{P}\left[\sum_{b=1}^{B} \sum_{i=1}^{n_B}\left\{ 2\sigma_e^2 + (\pi_j - \pi_{j-1} + \tau_{\text{d}_B{(j,b)}} - \tau_{\text{d}_B{(j-1,b)}})^2\right\} \right.\\ & \qquad \qquad \left. - n_B\sum_{b=1}^{B}\left\{\frac{2\sigma_e^2}{n_B} + (\pi_j - \pi_{j-1} + \tau_{\text{d}_B{(j,b)}} - \tau_{\text{d}_B{(j-1,b)}})^2 \right\} \right],\\
&= \sum_{j=2}^{P}\left[ 2(n_{\text{int}} - B)\sigma_e^2 + \sum_{b=1}^{B} \sum_{i=1}^{n_B}(\pi_j - \pi_{j-1} + \tau_{\text{d}_B{(j,b)}} - \tau_{\text{d}_B{(j-1,b)}})^2 \right.\\
& \qquad \qquad \left. \phantom{\sum_{j=2}^{D}} - n_B\sum_{b=1}^{B}(\pi_j - \pi_{j-1} + \tau_{\text{d}_B{(j,b)}} - \tau_{\text{d}_B{(j-1,b)}})^2\right],\\
&= \sum_{j=2}^{P}\left[2(n_{\text{int}} - B)\sigma_e^2\right],\\
&= 2(P - 1)(n_{\text{int}} - B)\sigma_e^2.
\end{align*}
Thus $\text{E}(\hat{\sigma}_e^2) = \sigma_e^2$, and the estimator is unbiased.

For non-complete-block treatment allocation, we can in this case take
	\begin{equation*}
	\hat{\sigma}_{b}^2 = \frac{1}{2}\left\{ \frac{1}{2(P - 1)(n_{\text{int}} - B)}\sum_{j=2}^{P}\sum_{b=1}^{B}\sum_{i=1}^{n_B}(q_{ijb} - \bar{q}_{jb})^2 - \hat{\sigma}_{e}^2 \right\}.
	\end{equation*}
	To show this is unbiased, again without loss of generality consider the case where $\mu_0=0$. Then
	\begin{align*}
	\text{E}(q_{ijb}) &= \pi_j + \pi_{j-1}+\tau_{\text{d}_B{(j,b)}}+\tau_{\text{d}_B{(j-1,b)}},\\
	\text{Var}(q_{ijb}) &= 2(\sigma_e^2+2\sigma_b^2),\\
	\text{E}(\bar{q}_{jb}) &= \pi_j + \pi_{j-1}+\tau_{\text{d}_B{(j,b)}}+\tau_{\text{d}_B{(j-1,b)}},\\
	\text{Var}(\bar{q}_{jb}) &= \frac{2(\sigma_e^2+2\sigma_b^2)}{n_B}.\\
	\end{align*}
	Using these results, a direct modification of the derivation for $\hat{\sigma}_{e}^2$ gives $\text{E}(\hat{\sigma}_{b}^2)=\sigma_b^2$.
	
	\section{Example 1: TOMADO}
	
	\subsection{Introduction}
	
	In this section, we expand on our results from Section 3 on Example 1; the TOMADO trial. Throughout, we set
	\[ \mu_0 = 10.65, \ \sigma_e^2=6.51, \ \alpha = 0.05, \ \beta = 0.2.\]
	However, we now consider, in turn, the effect of altering the values of $\sigma_b^2$, $\delta$, and the $\pi_j$, from those used in Section 3.
	
	\subsection{Influence of $\sigma_b^2$}
	
	First, we examine, in the case where $\pi_2 = -0.77$, $\pi_3 = -0.96$, $\pi_4 = -0.55$, and $\delta=-1.24$, the influence of the value of $\sigma_b^2$ on the performance of the re-estimation procedures. As in Section 3.5, we would like to ascertain the effect changing $\sigma_b^2$ has upon the FWER and power.
	
	Figures 6 and 7 respectively present our results on the FWER and power of the various re-estimation procedures when $n_{\text{int}}\in\{16,32\}$ for several values of $\sigma_b^2\in[0.25(10.12),4(10.12)]$, under the global null and alternative hypotheses respectively.
	
	As would be anticipated, the value of $\sigma_b^2$ appears to have extremely little effect upon the performance of each of the re-estimation procedures. Recall that this is a consequence of the fact that the complete block sequences utilised in Example 1 render the requisite sample size independent of the between person variance.
	
	\begin{figure}[htb]
		\begin{center}\label{TDS1_sigma_b_FWER}
			\includegraphics[width = 14cm]{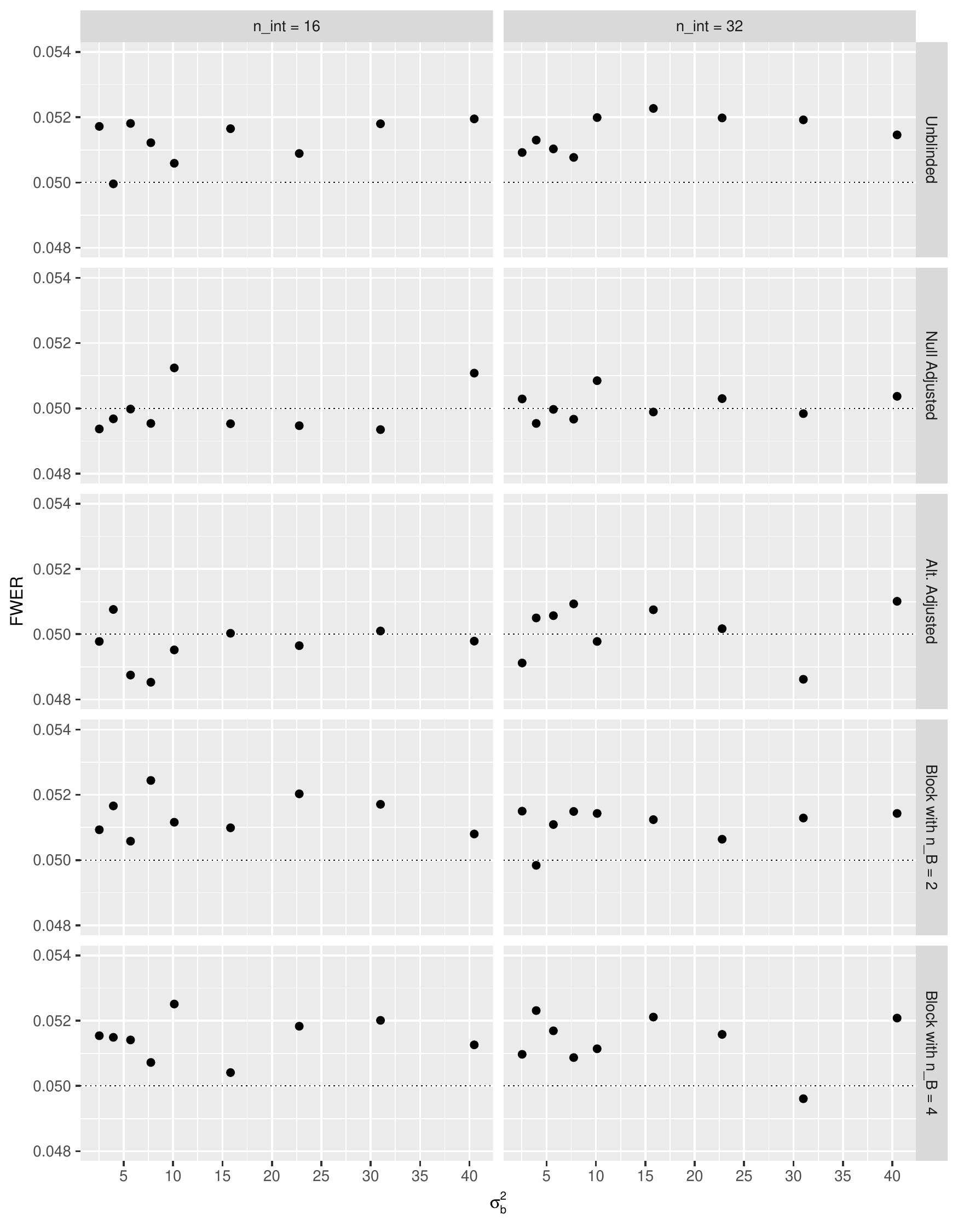}
			\caption{The simulated familywise error-rate (FWER) is shown under the global null hypothesis for each of the re-estimation procedures when $n_{\text{int}}\in\{16,32\}$, as a function of the between person variance $\sigma_b^2$, for Example 1. The Monte Carlo error is approximately 0.0007 in each instance. The dashed line indicates the desired value of the FWER.}		
		\end{center}
	\end{figure}
	
	\begin{figure}[htb]
		\begin{center}\label{TDS1_sigma_b_power}
			\includegraphics[width = 14cm]{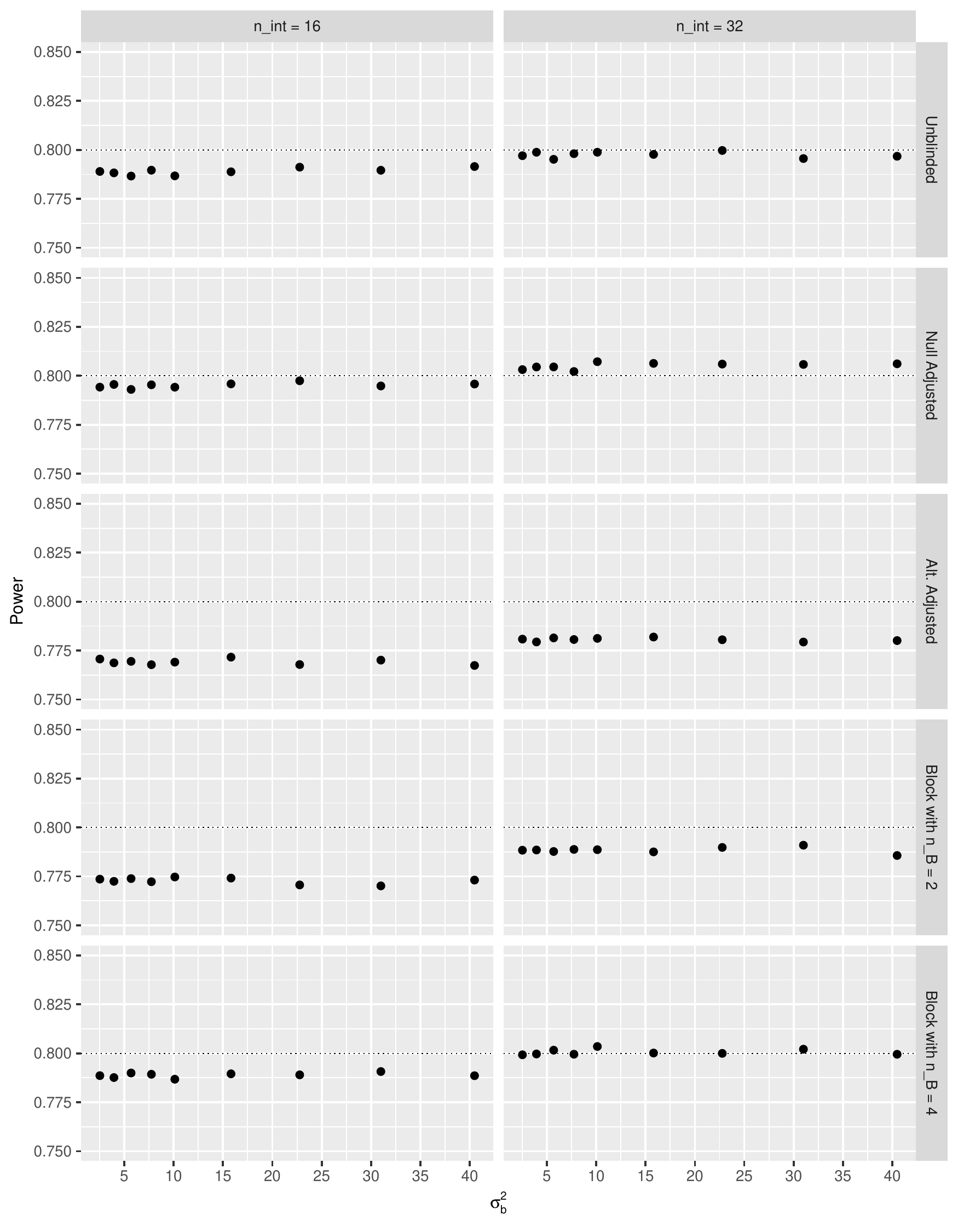}
			\caption{The simulated power is shown under the global alternative hypothesis for each of the re-estimation procedures when $n_{\text{int}}\in\{16,32\}$, as a function of the within person variance $\sigma_b^2$, for Example 1. The Monte Carlo error is approximately 0.0013 in each instance. The dashed line indicates the desired value of the power.}
		\end{center}
	\end{figure}
	
	\subsection{Influence of $\delta$}
	
	Here, we consider the case where $\pi_2 = -0.77$, $\pi_3 = -0.96$, $\pi_4 = -0.55$, and $\sigma_b^2=10.12$, focusing on the influence $\delta$ has upon the procedures FWER and power. Precisely, Figures 8 and 9 respectively present our findings for the FWER and power of the various re-estimation procedures when $n_{\text{int}}\in\{16,32\}$ for several values of $\delta\in[2(-1.24),0.5(-1.24)]$, under the global null and alternative hypotheses respectively.
	
	In Figure 8 we can see that there is no clear pattern to the effect on the FWER of changing $\delta$, with the fluctuations for several of the estimators relatively small. However, there is some evidence to suggest that increasing the value of $\delta$ (that is, making it closer to zero) reduces the FWER, as may be expected as this implies a larger requisite sample size.
	
	Similar statements are true for the power when examining Figure 9. Analogous to our discussions around Figure 4, the re-estimation procedures are over-powered when $n_{\text{int}}=32$ and $\delta$ is large in magnitude. Furthermore, increasing the value of $n_{\text{int}}$ once more universally increases the power, whilst there appears to be a point beyond which the power remains relatively constant.
	
	\begin{figure}[htb]
		\begin{center}\label{TDS1_delta_null}
			\includegraphics[width = 15cm]{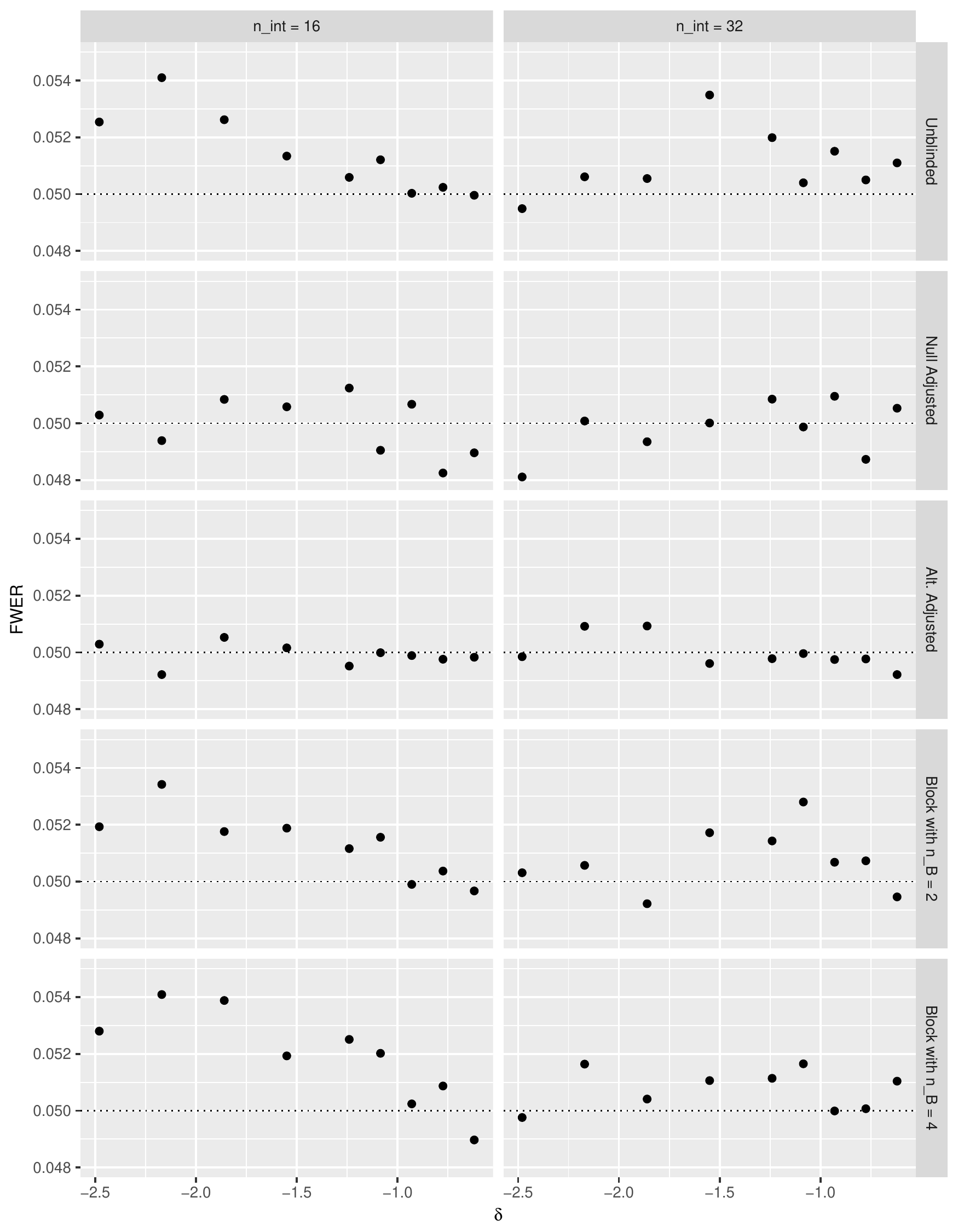}
			\caption{The simulated familywise error-rate (FWER) is shown under the global null hypothesis for each of the re-estimation procedures when $n_{\text{int}}\in\{16,32\}$, as a function of the clinically relevant difference $\delta$, for Example 1. The Monte Carlo error is approximately 0.0007 in each instance. The dashed line indicates the desired value of the FWER.}		
		\end{center}
	\end{figure}
	
	\begin{figure}[htb]
		\begin{center}\label{TDS1_delta_alt}
			\includegraphics[width = 15cm]{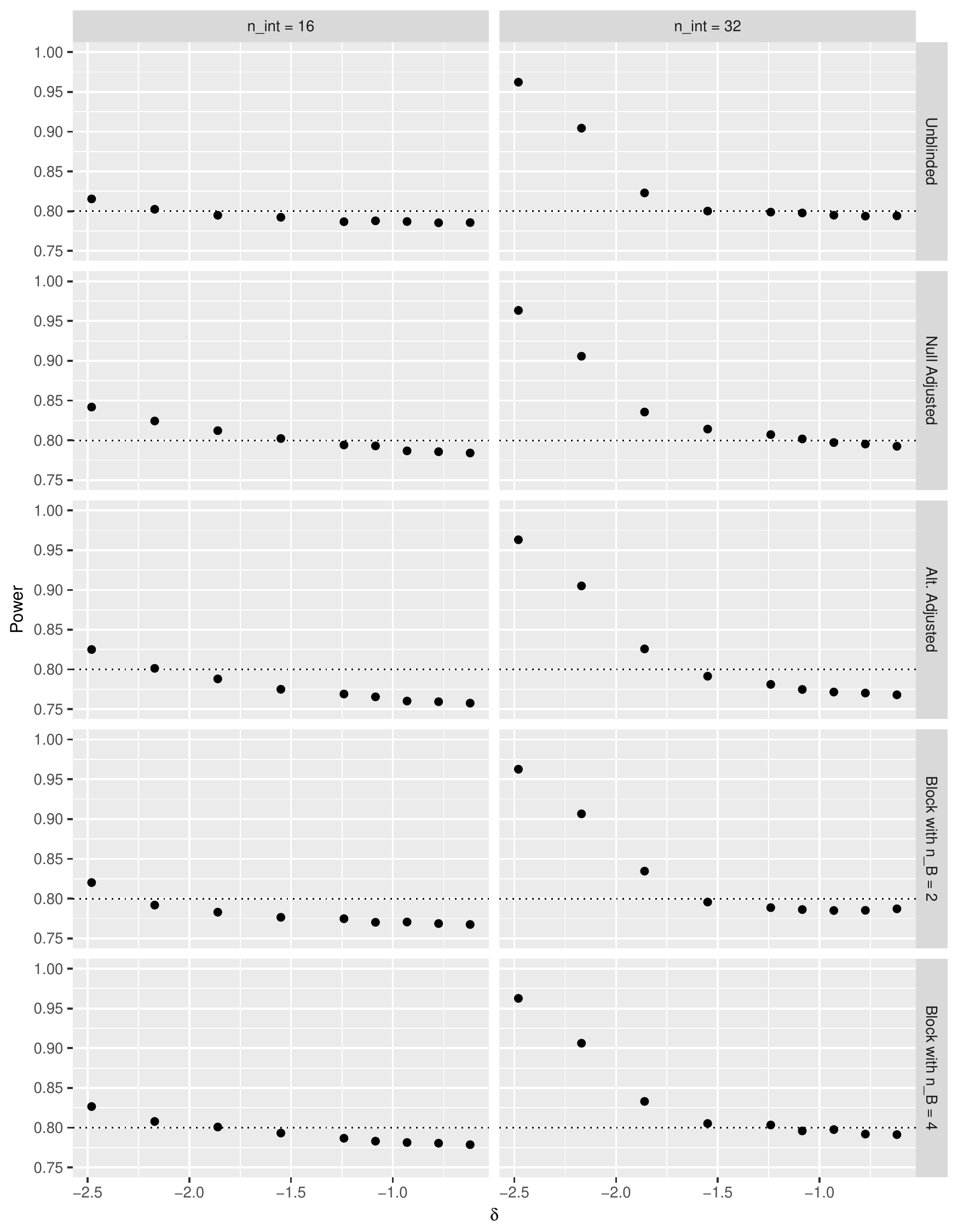}
			\caption{The simulated power is shown under the global alternative hypothesis for each of the re-estimation procedures when $n_{\text{int}}\in\{16,32\}$, as a function of the clinically relevant difference $\delta$, for Example 1. The Monte Carlo error is approximately 0.0013 in each instance. The dashed line indicates the desired value of the power.}
		\end{center}
	\end{figure}
	
	\subsection{Influence of period effects}
	
	In Section 4, we discussed how the influence of the period effects should be examined using simulation. Again, though the final analysis on the trials data is asymptotically invariant to the true value of the period effects, their value may influence the ability of the various re-estimation procedures to accurately estimate the variance parameters at the interim reassessment. In this section, we assess the influence of the value of the period effects on each of the re-estimation procedures. Explicitly, we consider the case with $n_{\text{int}}=16$ and $\sigma_e^2=6.51$, under the global null hypothesis. All others parameters are left as specified in Section 3.2, except for the $\pi_j$ for $j=2,3,4$. For these, in each replicate simulation we take
	$$ \pi_j \sim N(\hat{\pi}_j,\sigma_{\pi}^2), \qquad j=2,3,4. $$
	
	That is, a value for each $\pi_j$ is drawn from a normal distribution with mean $\hat{\pi}_j$, and specified variance $\sigma_{\pi}^2$. We then assess the effect of the size of $\sigma_{\pi}^2$ upon the trials FWER and values for $\hat{\sigma}_{e}^2$. Figures 10 and 11 display our findings. From them, it does appear that at least for this design scenario and associated parameter set, the value of the period effects tends to have little influence on the interim estimation of the within person variance, or on the trials FWER.
	
	\begin{figure}[htb]
		\begin{center}\label{period1}
			\includegraphics[width=16cm]{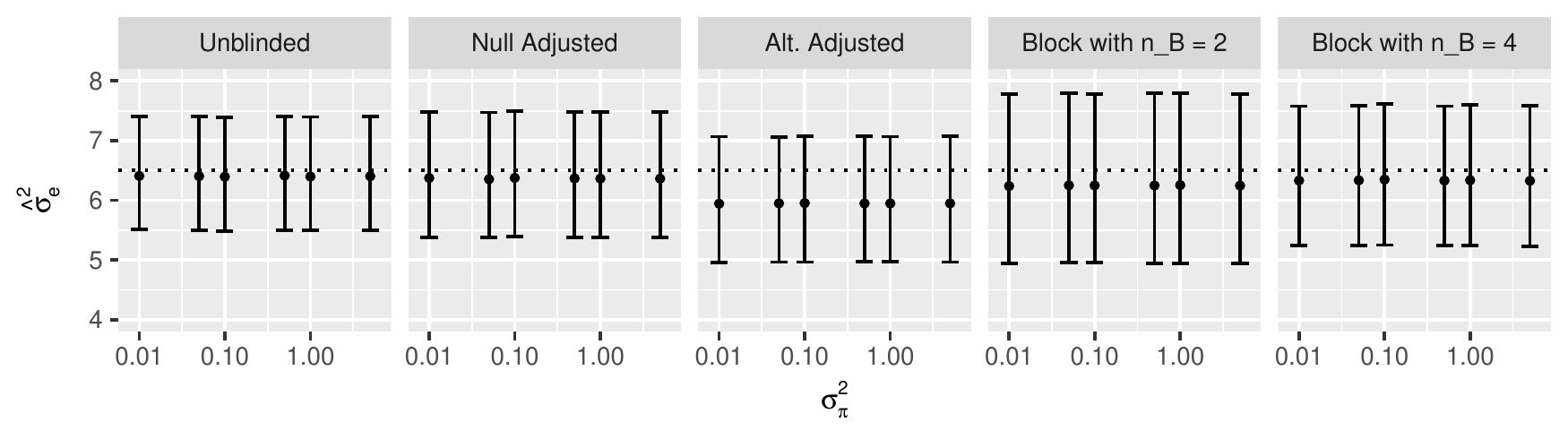}
			\caption{The distribution of $\hat{\sigma}_e^2$ is shown under the global null hypothesis for each of the re-estimation procedures  when $n_{\text{int}}=16$, as a function of the variance of the period effects $\sigma_{\pi}^2$, for Example 1. Precisely, for each scenario, the median, lower and upper quartile values of $\hat{\sigma}_e^2$ across the simulations are given. The dashed line indicates the true value of $\sigma_e^2$.}
		\end{center}
	\end{figure}
	
	\begin{figure}[htb]
		\begin{center}\label{period2}
			\includegraphics[width=16cm]{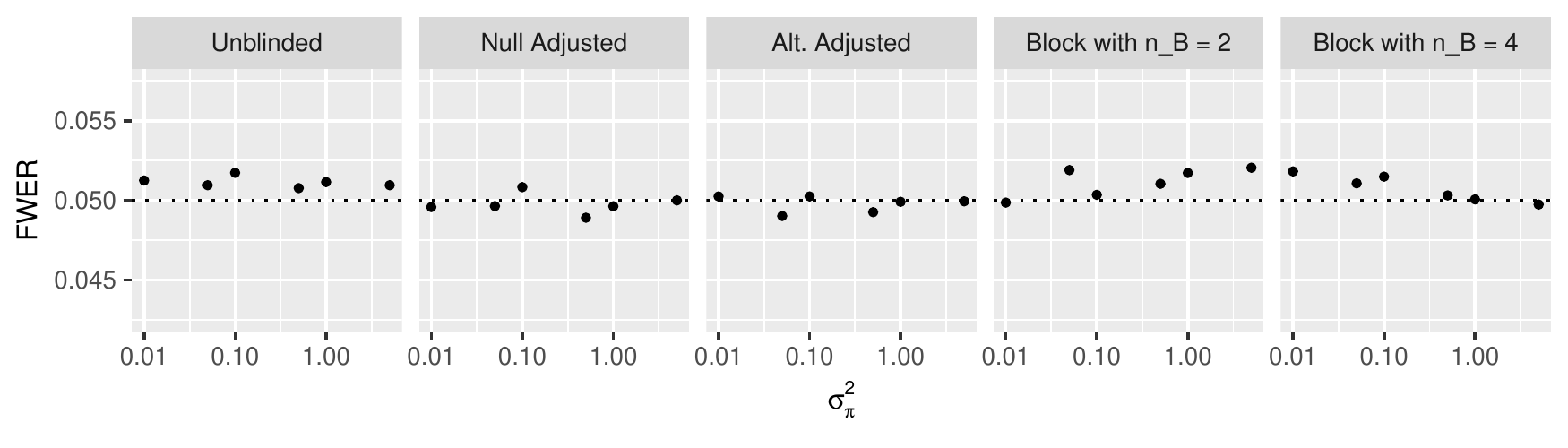}
			\caption{The simulated familywise error-rate (FWER) is shown under the global null hypothesis for each of the re-estimation procedures when $n_{\text{int}}=16$, as a function of the variance of the period effects $\sigma_{\pi}^2$, for Example 1. The Monte Carlo error is approximately 0.0007 in each instance. The dashed line indicates the desired value of the FWER.}
		\end{center}
	\end{figure}
	
	\section{Example 2: Formoterol}
		
		\subsection{Introduction}
		
		Senn (2002) reported the results of a double-blind placebo controlled cross-over trial to assess the performance of two doses of formoterol solution aerosol ($D=3$). Denoting the treatments by $d=0,1,2$, with $d=0$ the placebo and $d=1,2$ corresponding to the two doses of formoterol, patients were allocated to one of the following six incomplete block sequences
		\[ 01,\ 10,\ 02,\ 20,\ 12,\ 21. \]
		We assume that the primary FEV1 (forced expiratory volume in 1 second) outcome data was to be analysed with the linear mixed model (1), in order to test the following hypotheses
		\[ H_{0d} : \tau_d \le 0,\qquad H_{1d} : \tau_d > 0,\qquad d=1,2.\]
		A complete case analysis of the data presented by Senn (2002) gives the following estimates for the parameters in the model
		\[ \hat{\mu}_0 = 1.51, \ \hat{\pi}_2 = 0.03, \ \hat{\tau}_1 = 0.50, \ \hat{\tau}_2 = 0.52, \ \hat{\sigma}_e^2 = 0.053, \ \hat{\sigma}_b^2 = 0.49. \]
		Consequently, for $\alpha=0.1$ and $\beta=0.2$, a sample size of 30 patients would be powered to detect a clinically relevant difference of $\delta=0.2$.
		
		Accordingly, in this section we investigate the performance of the re-estimation procedures with the various design parameters motivated by those obtained from the results of this formoterol trial. In particular, this allows us to consider performance in a more challenging small sample setting. In all of what follows we set $\mu_0=1.51$, $\pi_2=0.03$, $\alpha=0.1$ and $\beta=0.2$, and we suppose that patients are allocated treatments via one of the six sequences listed above ($P=2$). We then consider in turn the effect of varying one of the parameters $\sigma_e^2$, $\sigma_b^2$, and $\delta$. As for Example 1, we will consider performance under the global null hypothesis ($\tau_1=\tau_2=0$), when only treatment one is effective ($\tau_1=\delta,\tau_2=0$), under the global alternative hypothesis ($\tau_1=\tau_2=\delta$), and under the observed treatment effects ($\tau_1=0.50, \tau_2=0.52$). Moreover, we again take $n_{\text{max}}=1000$, and estimate the average performance of each design and analysis procedure using 100,000 trial simulations.
		
		\subsection{Distributions of $\hat{\sigma}_e^2$, $\hat{\sigma}_b^2$ and $\hat{N}$}
		
		We first examine the performance of the re-estimation procedures when $\sigma_e^2=0.053$, $\sigma_b^2=0.49$, and $\delta=0.2$, for $n_{\text{int}}=18$. The resulting distributions of $\hat{\sigma}_e^2$, $\hat{\sigma}_b^2$, and $\hat{N}$, are shown in Figures 12-14, via their median, lower and upper quartiles across the simulations. The results are grouped according to the true value of the treatment effects.
		
		Following our findings for Example 1, the median values of $\hat{\sigma}_e^2$, $\hat{\sigma}_b^2$, and $\hat{N}$ for the unblinded and block randomised procedures are always close to their respective true values. However, this is not always the case for the adjusted estimators. In particular, the null adjusted estimator over-estimates the value of $\sigma_e^2$ when $\tau_1=\delta$, and the alternative adjusted procedure under estimates $\sigma_e^2$ when $\tau_1=\tau_2=0$. Both adjusted estimators perform extremely badly at re-estimating $\sigma_e^2$ under the observed treatment effects.
		
		In terms of $\hat{\sigma}_b^2$, the performance of the re-estimation procedures is more comparable, though the adjusted procedures fair worse under the observed treatment effects.
		
		As expected, the results for $\hat{N}$ once again mirror those for $\hat{\sigma}_e^2$. This means, in particular, that the median value of $\hat{N}$ is substantially larger for the adjusted estimators under the observed treatment effects, and is slightly larger for the null adjusted procedure when $\tau_1=\delta$.
		
		Increasing the value of $n_{\text{int}}$ reduces the interquartile range for $\hat{\sigma}_e^2$ and $\hat{N}$ for each procedure, and results in median values closer to the truth, as would be expected. Finally, we observe that the interquartile range for the unblinded procedure is often smaller than that of its adjusted or block randomisation counterparts.
		
		\begin{figure}[htb]
			\begin{center}\label{TDS2_sigmae2hat}
				\includegraphics[width = 15cm]{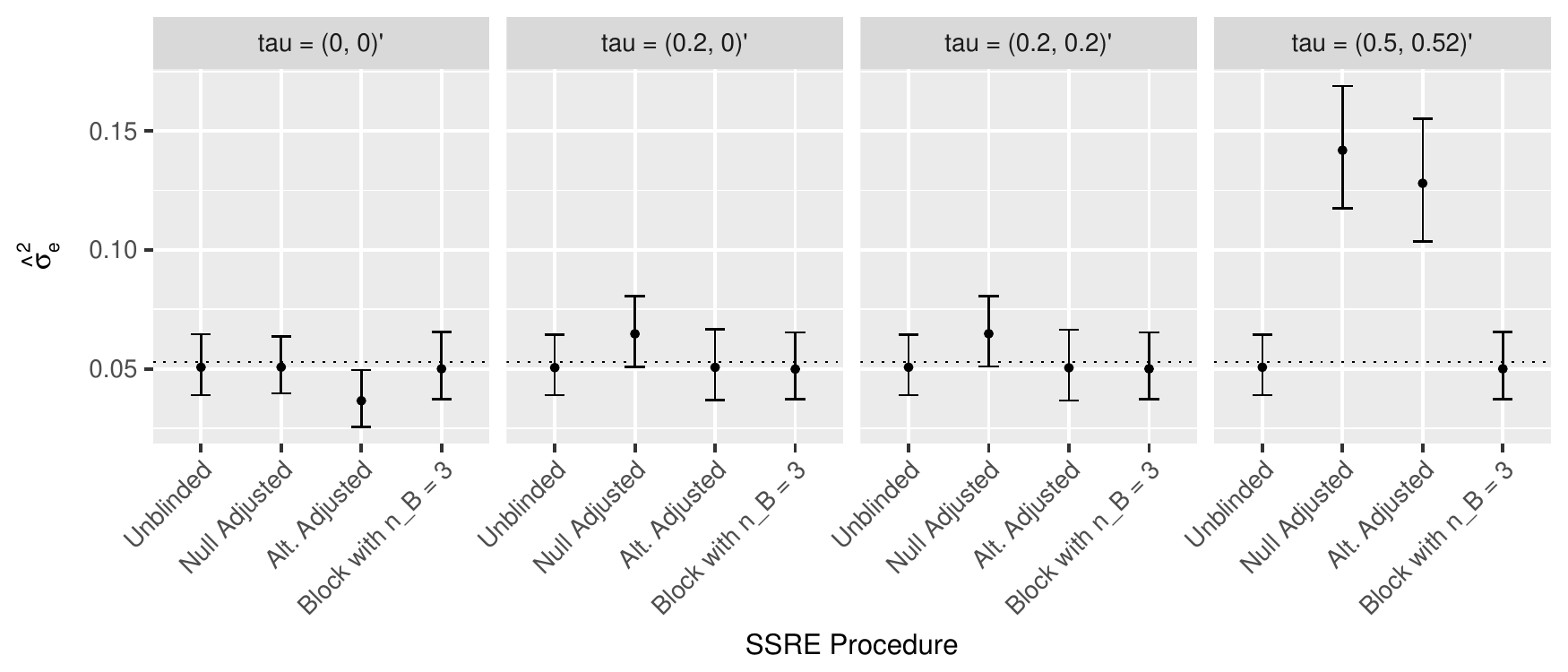}
				\caption{The distribution of $\hat{\sigma}_e^2$ is shown for each of the re-estimation procedures for several values of $\boldsymbol{\tau}$, for Example 2. Precisely, for each scenario, the median, lower and upper quartile values of $\hat{\sigma}_e^2$ across the simulations are given. The dashed line indicates the true value of $\sigma_e^2$.}
			\end{center}
		\end{figure}
		
		\begin{figure}[htb]
			\begin{center}\label{TDS2_sigmab2hat}
				\includegraphics[width = 15cm]{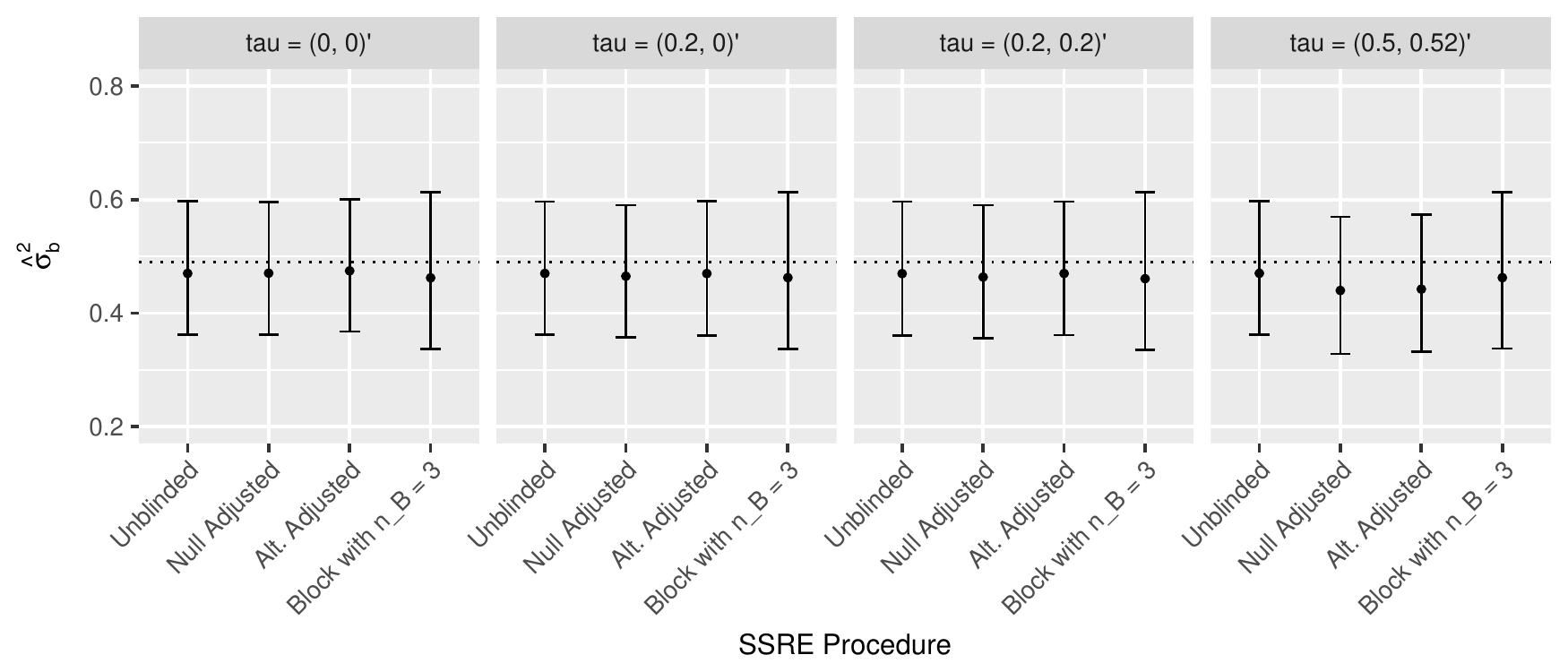}
				\caption{The distribution of $\hat{\sigma}_b^2$ is shown for each of the re-estimation procedures for several values of $\boldsymbol{\tau}$, for Example 2. Precisely, for each scenario, the median, lower and upper quartile values of $\hat{\sigma}_b^2$ across the simulations are given. The dashed line indicates the true value of $\sigma_b^2$.}
			\end{center}
		\end{figure}
		
		\begin{figure}[htb]
			\begin{center}\label{TDS2_Nhat}
				\includegraphics[width = 15cm]{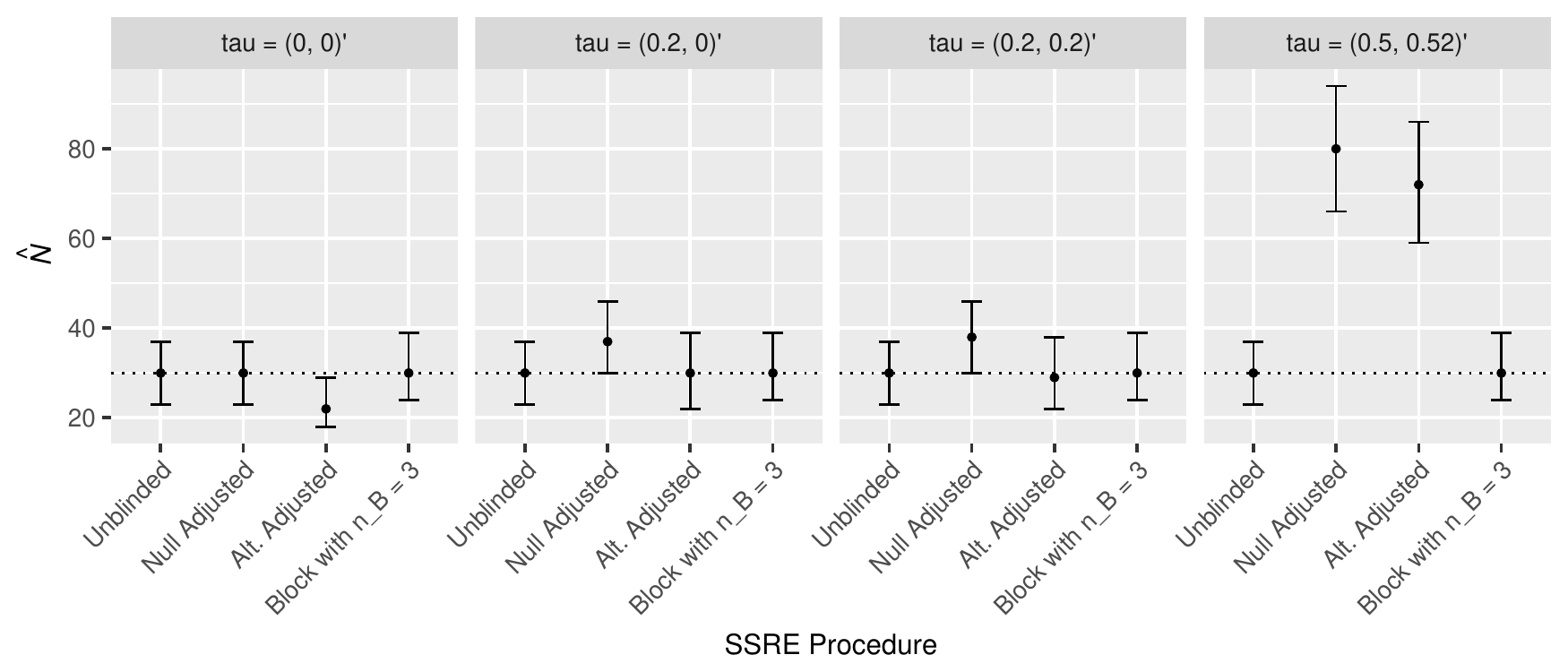}
				\caption{The distribution of $\hat{N}$ is shown for each of the re-estimation procedures for several values of $\boldsymbol{\tau}$, for Example 2. Precisely, for each scenario, the median, lower and upper quartile values of $\hat{N}$ across the simulations are given. The dashed line indicates the true required value of $N$.}
			\end{center}
		\end{figure}
		
		\subsection{Familywise error-rate, power, and sample size inflation factor}
		
		For the scenarios from Section S.M.5.2 that were not conducted under the observed treatment effects, the estimated FWER and power were also recorded. Additional simulations were also performed to ascertain the power of the procedures under the global alternative hypothesis when the sample size inflation factor introduced in Section 3.6 was utilised. The results are displayed in Table 2.
		
		We can see that the unblinded and block randomised procedures experience similar and substantial inflation to the FWER. In contrast the adjusted procedures have an equal FWER with much smaller inflation above the nominal level. Thus for this example, the small requisite sample size appears to have inhibited the ability of the re-estimation procedures to retain a FWER close to $\alpha$.
		
		Each of the procedures attain the desired power apart from that utilising the alternative adjusted estimator. The null adjusted estimator provides the largest power. These findings are not surprising in light of the distributions of $\hat{\sigma}_e^2$ observed in Figure 12. Therefore in this instance the inflation factor appears to only be of use to the alternative adjusted procedure.
		
		\subsection{Influence of $\sigma_e^2$}
		
		We now consider cases where $\sigma_b^2=0.49$ and $\delta=0.2$, but $\sigma_e^2\neq0.053$. Specifically, in Figures 15 and 16 we respectively examine the FWER and power of the re-estimation procedures under the global null and alternative hypotheses when $\sigma_e^2\in[0.25(0.053),4(0.053)]$.
		
		Figure 16 implies that the results for Example 2 are similar to those observed for Example 1. Explicitly, when $\sigma_e^2$ is very small the procedures are each over-powered, and as $\sigma_e^2$ increases there appears to be less of an effect upon the power.
		
		In contrast, the results on the FWER displayed in Figure 15 depict a distinctive pattern for each of the re-estimation procedures. Most likely, the observed peaks reflect a point at which the designs begin to switch from terminating the trial at the interim reassessment, to continuing to the end of the second stage. Overall, it is clear that the adjusted procedures typically have similar values of the FWER, with the same true of the unblinded and block randomised procedures. Moreover, whilst all of the estimators experience notable inflation to the FWER, it is smaller for the adjusted estimators, particularly as $\sigma_e^2$ increases.
		
		\begin{figure}[htb]
			\begin{center}\label{TDS2_sigma_e_FWER}
				\includegraphics[width = 15cm]{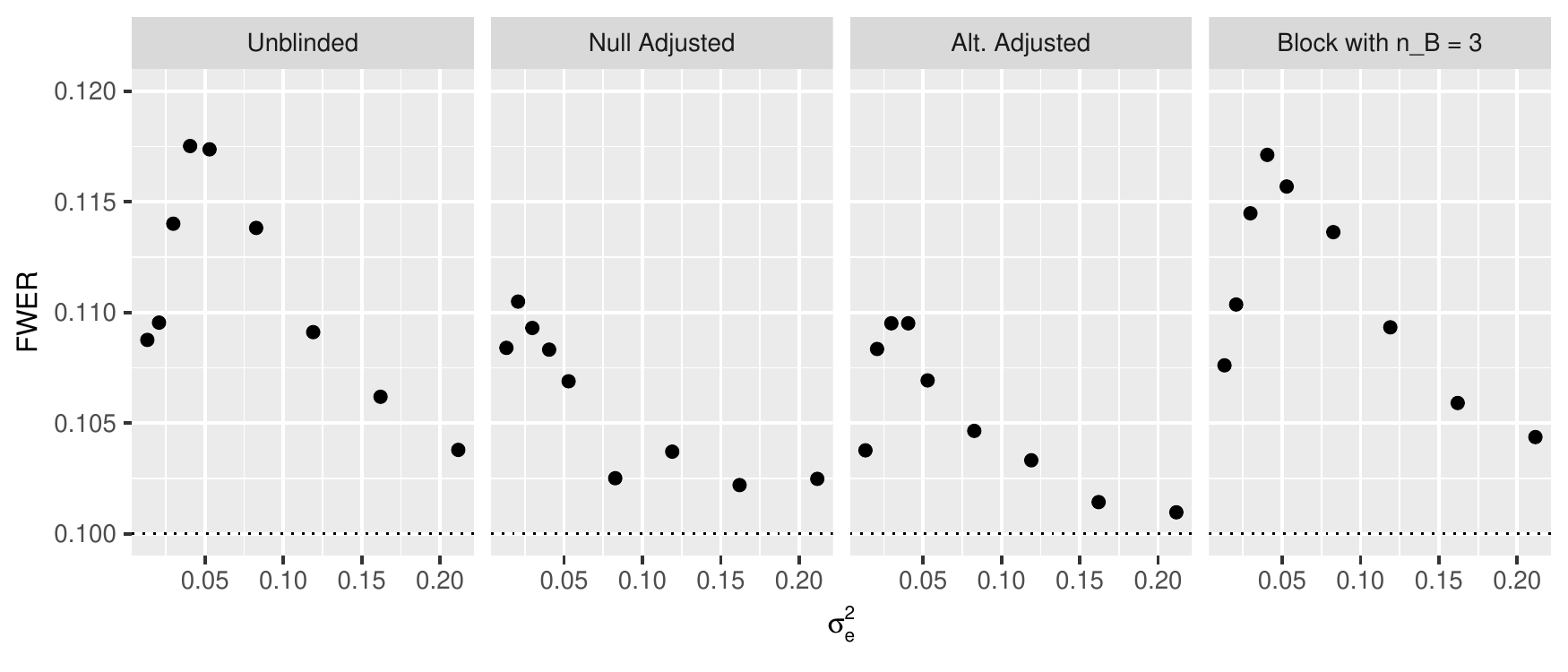}
				\caption{The simulated familywise error-rate (FWER) is shown under the global null hypothesis for each of the re-estimation procedures, as a function of the within person variance $\sigma_e^2$, for Example 2. The Monte Carlo error is approximately 0.001 in each instance. The dashed line indicates the desired value of the FWER.}		
			\end{center}
		\end{figure}
		
		\begin{figure}[htb]
			\begin{center}\label{TDS2_sigma_e_power}
				\includegraphics[width = 15cm]{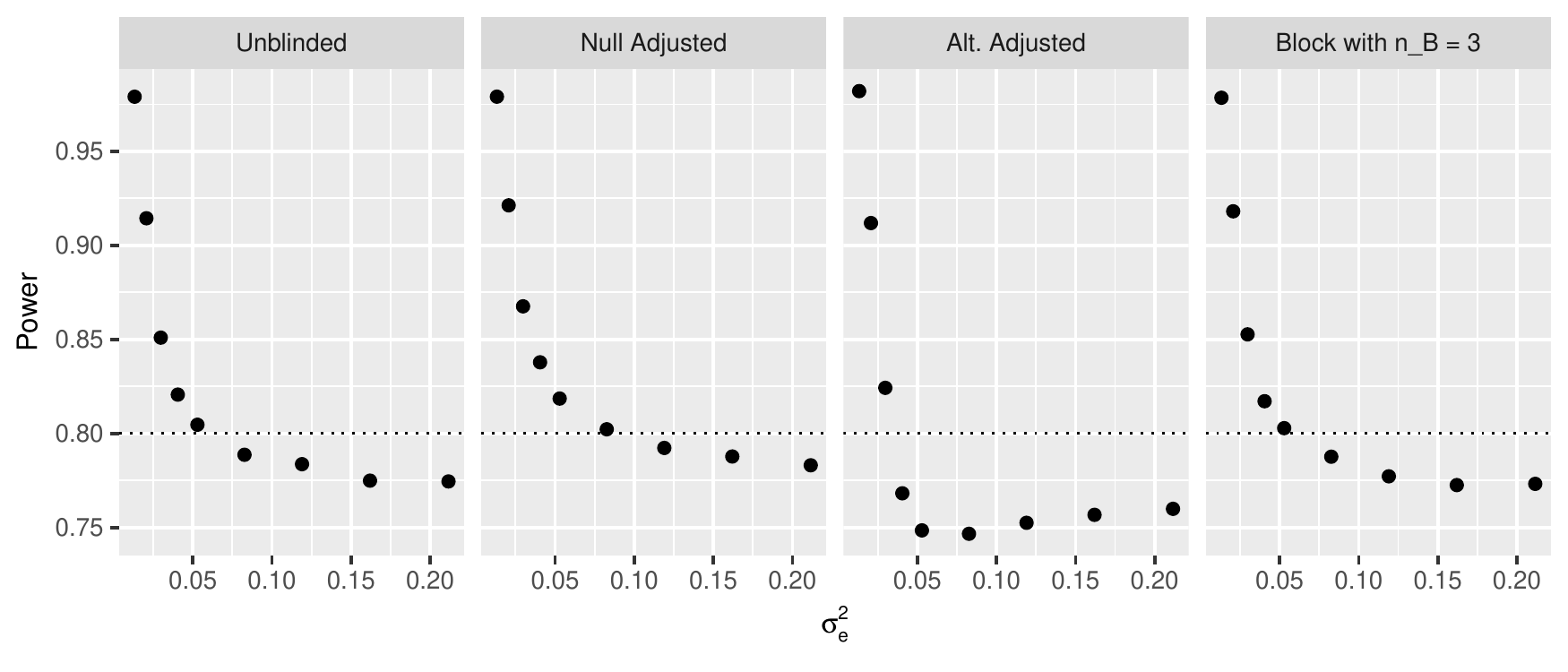}
				\caption{The simulated power is shown under the global alternative hypothesis for each of the re-estimation procedures, as a function of the within person variance $\sigma_e^2$, for Example 2. The Monte Carlo error is approximately 0.0013 in each instance. The dashed line indicates the desired value of the power.}
			\end{center}
		\end{figure}
		
		\subsection{Influence of $\sigma_b^2$}
		
		Next, we examine scenarios where $\sigma_e^2=0.053$ and $\delta=0.2$, but $\sigma_b^2\neq0.49$: in Figures 17 and 18 we respectively examine the FWER and power of the re-estimation procedures under the global null and alternative hypotheses when $\sigma_b^2\in[0.25(0.49),4(0.49)]$.
		
		Allowing for Monte Carlo error, it appears that the value of $\sigma_b^2$, as in Example 1, has negligible effect upon the FWER and power of each of the the re-estimation procedures. This may seem surprising as we specifically noted that a value for $\sigma_b^2$ would be required to estimate the required sample size of a trial using the incomplete block sequences specified in Section S.M.5.1. However, this result reflects two factors. The first is that, as evidenced by Figure 13, the re-estimation procedures are very effective at estimating the value of $\sigma_b^2$. More importantly, though, is that whilst the sample size required by the formoterol trial will be dependent upon $\sigma_b^2$, it will still be principally driven by the value of $\sigma_e^2$. Therefore, the main driver of the utility of the re-estimation procedures in this setting remains their ability to re-estimate the within person variance.
		
		\begin{figure}[htb]
			\begin{center}\label{TDS2_sigma_b_FWER}
				\includegraphics[width = 15cm]{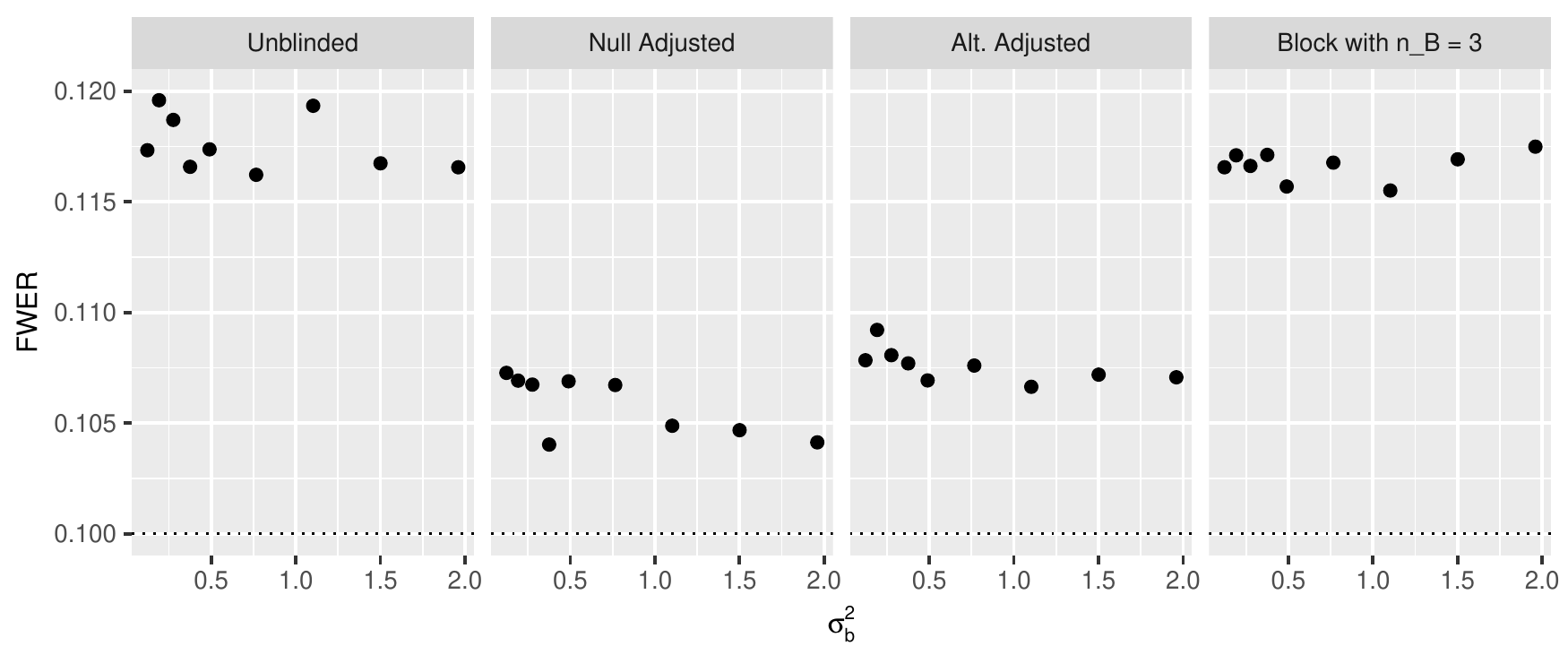}
				\caption{The simulated familywise error-rate (FWER) is shown under the global null hypothesis for each of the re-estimation procedures, as a function of the within person variance $\sigma_b^2$, for Example 2. The Monte Carlo error is approximately 0.001 in each instance. The dashed line indicates the desired value of the FWER.}		
			\end{center}
		\end{figure}
		
		\begin{figure}[htb]
			\begin{center}\label{TDS2_sigma_b_power}
				\includegraphics[width = 15cm]{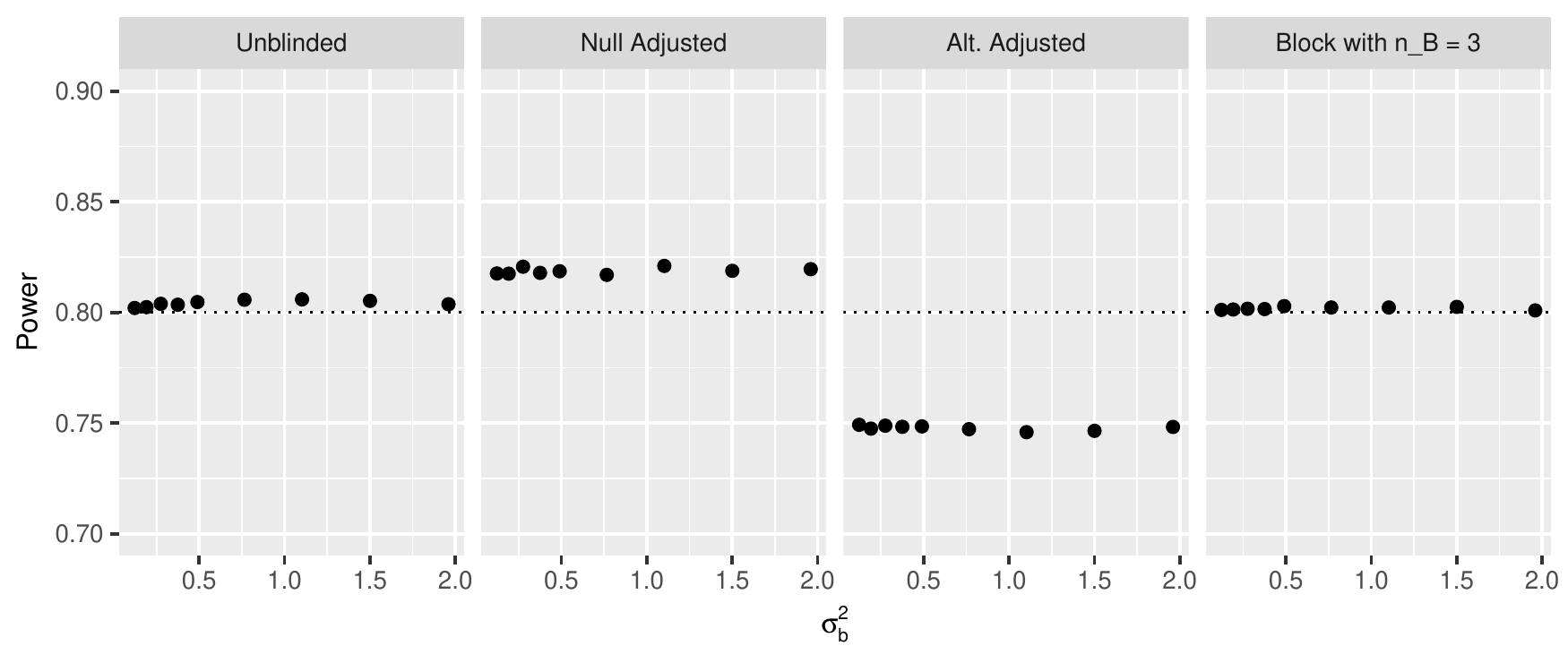}
				\caption{The simulated power is shown under the global alternative hypothesis for each of the re-estimation procedures, as a function of the within person variance $\sigma_b^2$, for Example 2. The Monte Carlo error is approximately 0.0013 in each instance. The dashed line indicates the desired value of the power.}
			\end{center}
		\end{figure}
		
		\subsection{Influence of $\delta$}
		
		In our final investigations for Example 2 we consider scenarios in which $\sigma_e^2=0.053$ and $\sigma_b^2=0.49$, but $\delta\neq0.2$. Precisely, in Figures 19 and 20 we respectively examine the FWER and power of the re-estimation procedures under the global null and alternative hypotheses when $\delta\in[0.5(0.2),2(0.2)]$.
		
		Examining Figure 19, the FWER for the unblinded and block randomised procedures display the same rising and falling shape as observed in Figure 15. In contrast, the FWER for the adjusted procedures appears only to rise with in $\delta$. This should not be surprising as larger values of $\delta$ imply smaller requisite sample sizes, leaving the procedures prone to small sample size issues.
		
		From Figure 20, as we would anticipate, for the larger considered values of $\delta$ the re-estimation procedures are substantially over-powered. All but the alternative adjusted procedure still perform well though when $\delta=0.1$.
		
		\begin{figure}[htb]
			\begin{center}\label{TDS2_delta_null}
				\includegraphics[width = 15cm]{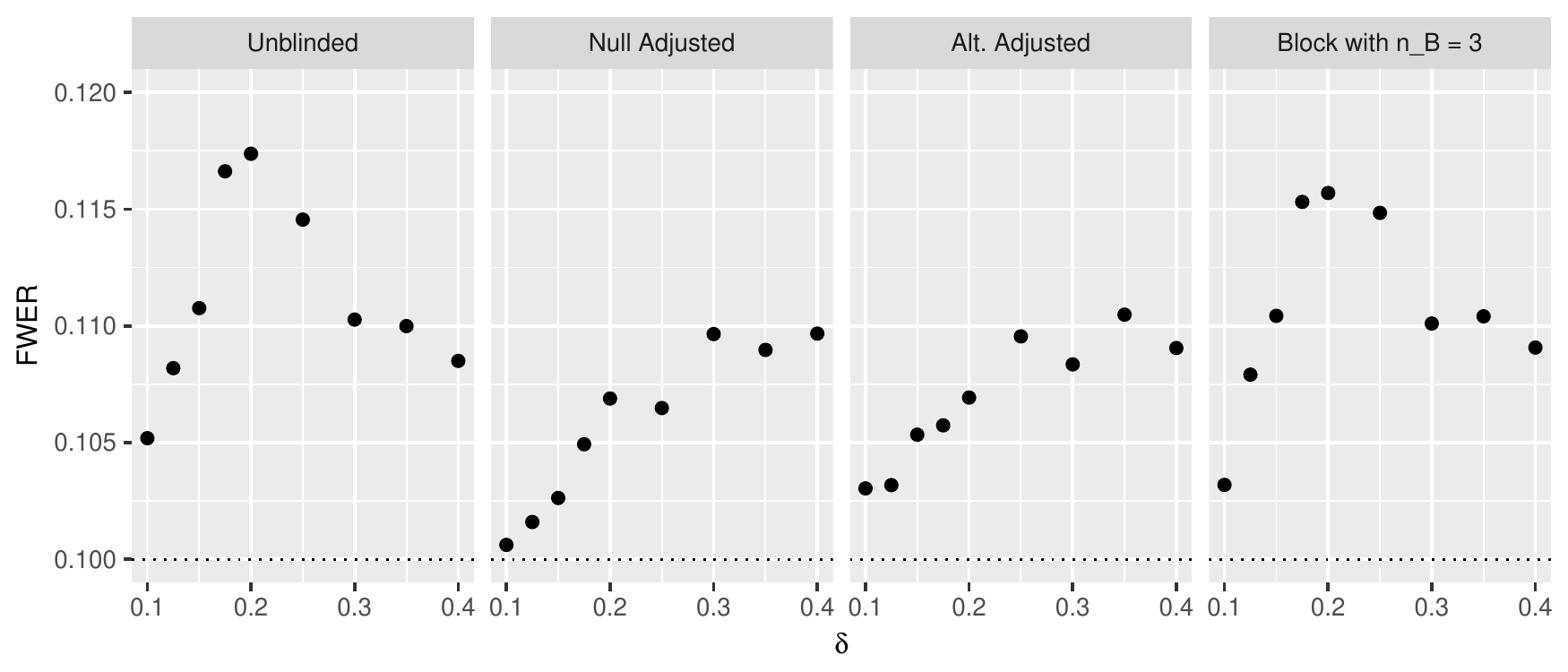}
				\caption{The simulated familywise error-rate (FWER) is shown under the global null hypothesis for each of the re-estimation procedures, as a function of the clinically relevant difference $\delta$, for Example 2. The Monte Carlo error is approximately 0.001 in each instance. The dashed line indicates the desired value of the FWER.}		
			\end{center}
		\end{figure}
		
		\begin{figure}[htb]
			\begin{center}\label{TDS2_delta_alt}
				\includegraphics[width = 15cm]{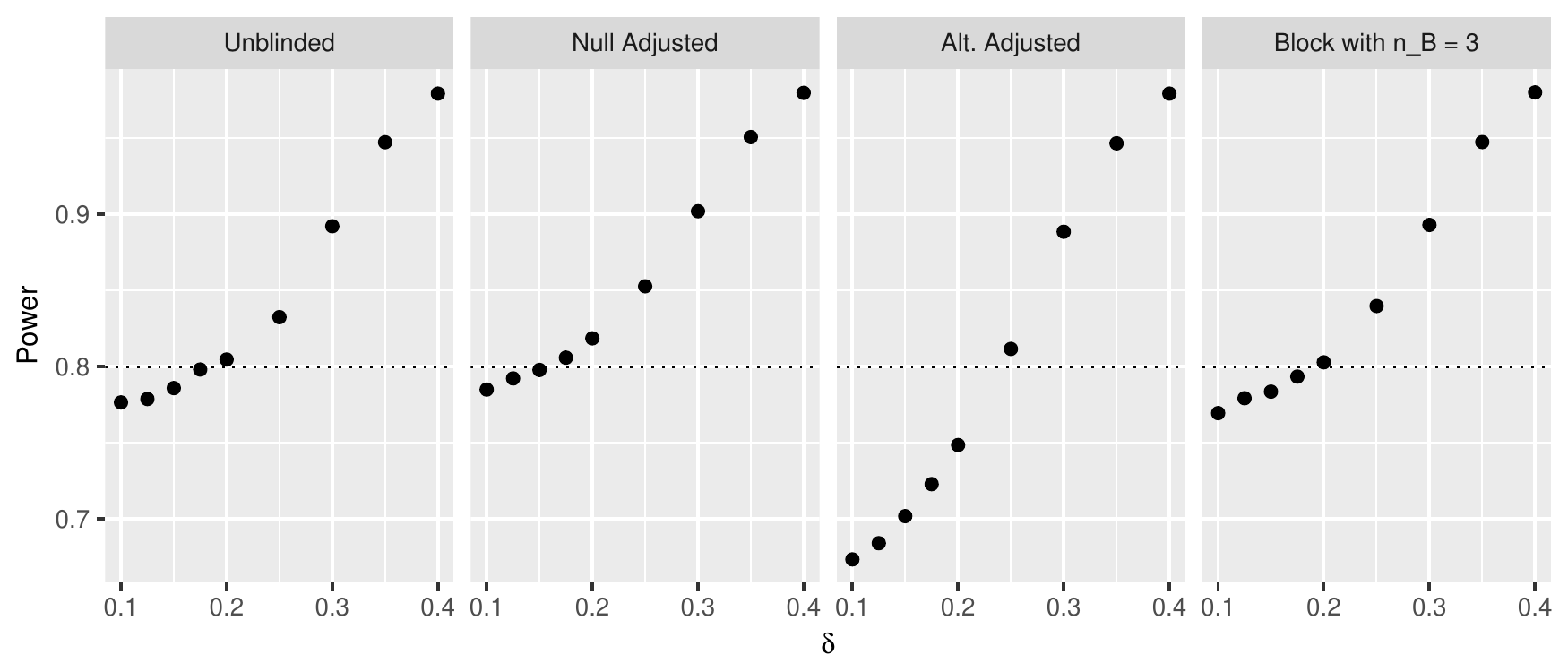}
				\caption{The simulated power is shown under the global alternative hypothesis for each of the re-estimation procedures, as a function of the clinically relevant difference $\delta$, for Example 2. The Monte Carlo error is approximately 0.0013 in each instance. The dashed line indicates the desired value of the power.}
			\end{center}
		\end{figure}
		
		\section{Example 3: Hypertension}
		
		\subsection{Introduction}
		
		Ebbutt (1984) presented an analysis of data from a two-treatment ($D=2$) three-period ($P=3$) crossover hypertension trial. Denoting the treatments by $d=0,1$, patients were assigned to one of the following four sequence groups
		\[ 011,\ 100,\ 010,\ 101. \]
		Whilst Ebbutt (1984) assessed the data from only ten subjects on each sequence, Jones and Kenward (2014) discussed the data from all patients who completed the trial. Here, we suppose that the data Jones and Kenward (2014) presented on the outcome systolic blood pressure were to be analysed using the linear mixed model (1) in order to test the following hypotheses
		\[ H_{01} : \tau_1 \ge 0,\qquad H_{11} : \tau_1 < 0.\]
		Thus, as in Example 1, a negative treatment effect implies efficacy.
		
		In analysing the data from this trial using linear mixed model (1) we find
		\[ \hat{\mu}_0 = 156.77, \ \hat{\pi}_2 = -2.13, \ \hat{\pi}_3=-4.90, \ \hat{\tau}_1 = -7.55, \ \hat{\sigma}_e^2 = 169.8, \ \hat{\sigma}_b^2 = 255.0. \]
		Therefore, for $\alpha=0.025$, the trial's sample size of 90 patients would provide power for a clinically relevant difference of $\delta=-5.39$ when $\beta=0.1$.
		
		In this section, we explore the performance of the re-estimation procedures with the parameters motivated by the results of this hypertension trial. Thus, having considered complete block sequences in Example 1, and incomplete block sequences in Example 2, we now consider a design utilising extra-period sequences.
		
		Following the same path as in Example 2, throughout what follows we set $\mu_0=156.77$, $\pi_2=-2.13$, $\pi_3=-4.90$, $\alpha=0.025$ and $\beta=0.1$, supposing that patients are allocated treatments based upon one of the four sequences given above. We then again consider the effect of varying the parameters $\sigma_e^2$, $\sigma_b^2$, and $\delta$. We examine performance under the null hypothesis ($\tau_1=0$), the alternative hypothesis ($\tau_1=\delta$), and under the observed treatment effect ($\tau_1=-7.55$). Finally, once more, we take $n_{\text{max}}=1000$, and estimate the average performance of each design and analysis procedure using 100,000 trial simulations.
		
		\subsection{Distributions of $\hat{\sigma}_e^2$, $\hat{\sigma}_b^2$ and $\hat{N}$}
		
		We first examine the performance of the re-estimation procedures when $\sigma_e^2=169.8$, $\sigma_b^2=255.0$, and $\delta=-5.39$, for $n_{\text{int}}\in\{16,32,48\}$. The resulting distributions of $\hat{\sigma}_e^2$, $\hat{\sigma}_b^2$, and $\hat{N}$, are shown in Figures 21-23, via their median, lower and upper quartiles across the simulations. The results are grouped according to the timing of the re-estimation and by the true value of the treatment effects.
		
		Overall, the results are very similar to those of Examples 1 and 2. In particular, the unblinded and block randomised procedures typically perform well in terms of re-estimating $\sigma_e^2$, whilst the adjusted estimators performance is highly dependent on the value of $\tau_1$.
		
		As in Example 1, increasing the value of $n_B$ for the block randomised procedure appears advantageous. Moreover, as expected, increasing the value of $n_{\text{int}}$ reduces the distance between the median values of $\hat{\sigma}_e^2$, $\hat{\sigma}_b^2$, and $\hat{N}$ and their respective true values, whilst also reducing the interquartile range.
		
		In this case, the distributions of the $\hat{\sigma}_b^2$ are comparable across the procedures, regardless of the value of $n_{\text{int}}$ or $\tau_1$.
		
		Once more, the results for $\hat{N}$ largely reflect those for $\hat{\sigma}_e^2$. However, the median values for $\hat{N}$ are larger when using the block randomised procedure with $n_B=8$. This is a consequence of the fact that by our assumptions an entire additional block of patients will be recruited even when a smaller number of patients are required to attain the requisite sample size.
		
		\begin{figure}[htb]
			\begin{center}\label{TDS3_sigmae2hat}
				\includegraphics[width = 15cm]{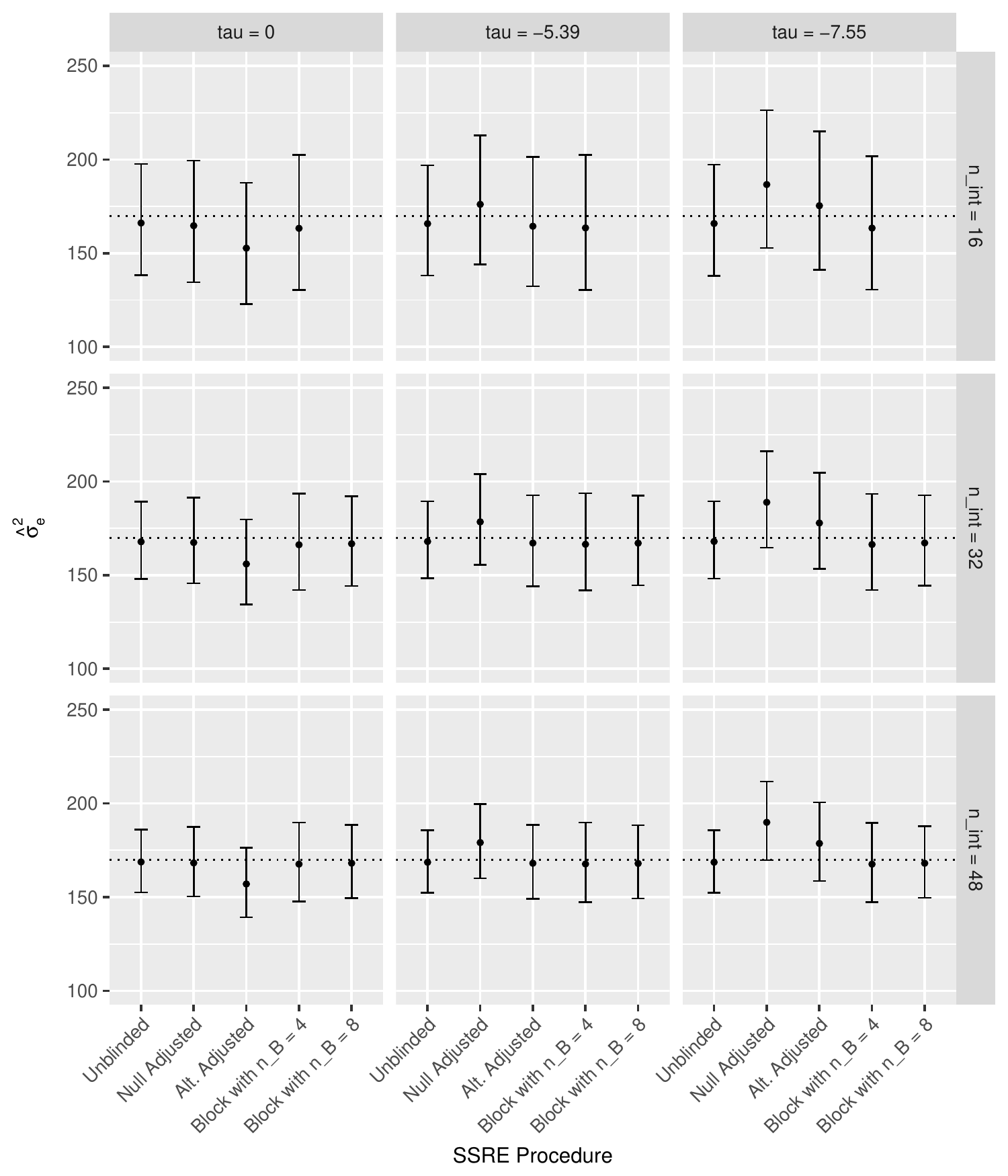}
				\caption{The distribution of $\hat{\sigma}_e^2$ is shown for each of the re-estimation procedures for several values of $\boldsymbol{\tau}$, and for several values of $n_{\text{int}}$, for Example 3. Precisely, for each scenario, the median, lower and upper quartile values of $\hat{\sigma}_e^2$ across the simulations are given. The dashed line indicates the true value of $\sigma_e^2$.}
			\end{center}
		\end{figure}
		
		\begin{figure}[htb]
			\begin{center}\label{TDS3_sigmab2hat}
				\includegraphics[width = 15cm]{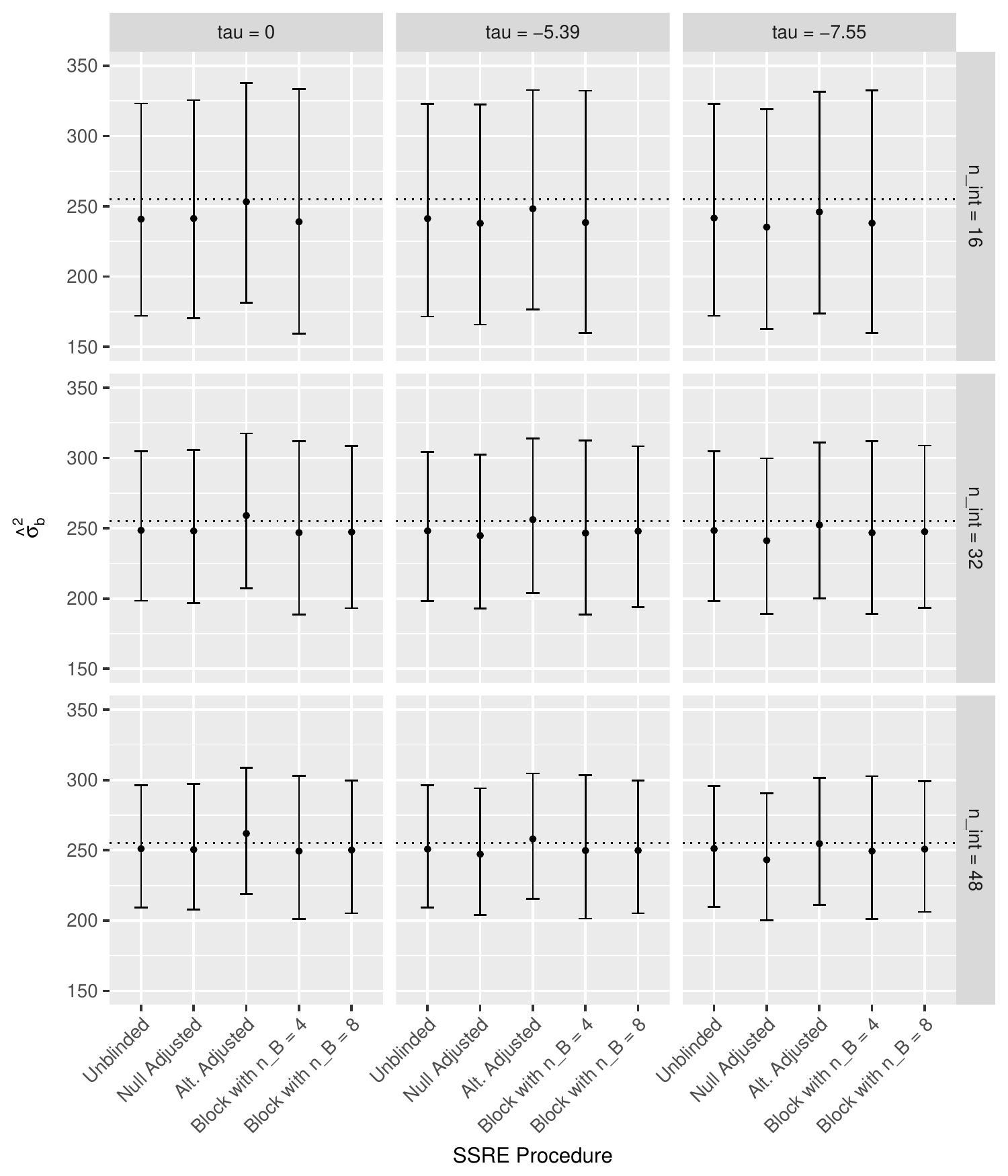}
				\caption{The distribution of $\hat{\sigma}_b^2$ is shown for each of the re-estimation procedures for several values of $\boldsymbol{\tau}$, and several values of $n_{\text{int}}$, for Example 3. Precisely, for each scenario, the median, lower and upper quartile values of $\hat{\sigma}_e^2$ across the simulations are given. The dashed line indicates the true value of $\sigma_e^2$.}
			\end{center}
		\end{figure}
		
		\begin{figure}[htb]
			\begin{center}\label{Nhat}
				\includegraphics[width = 15cm]{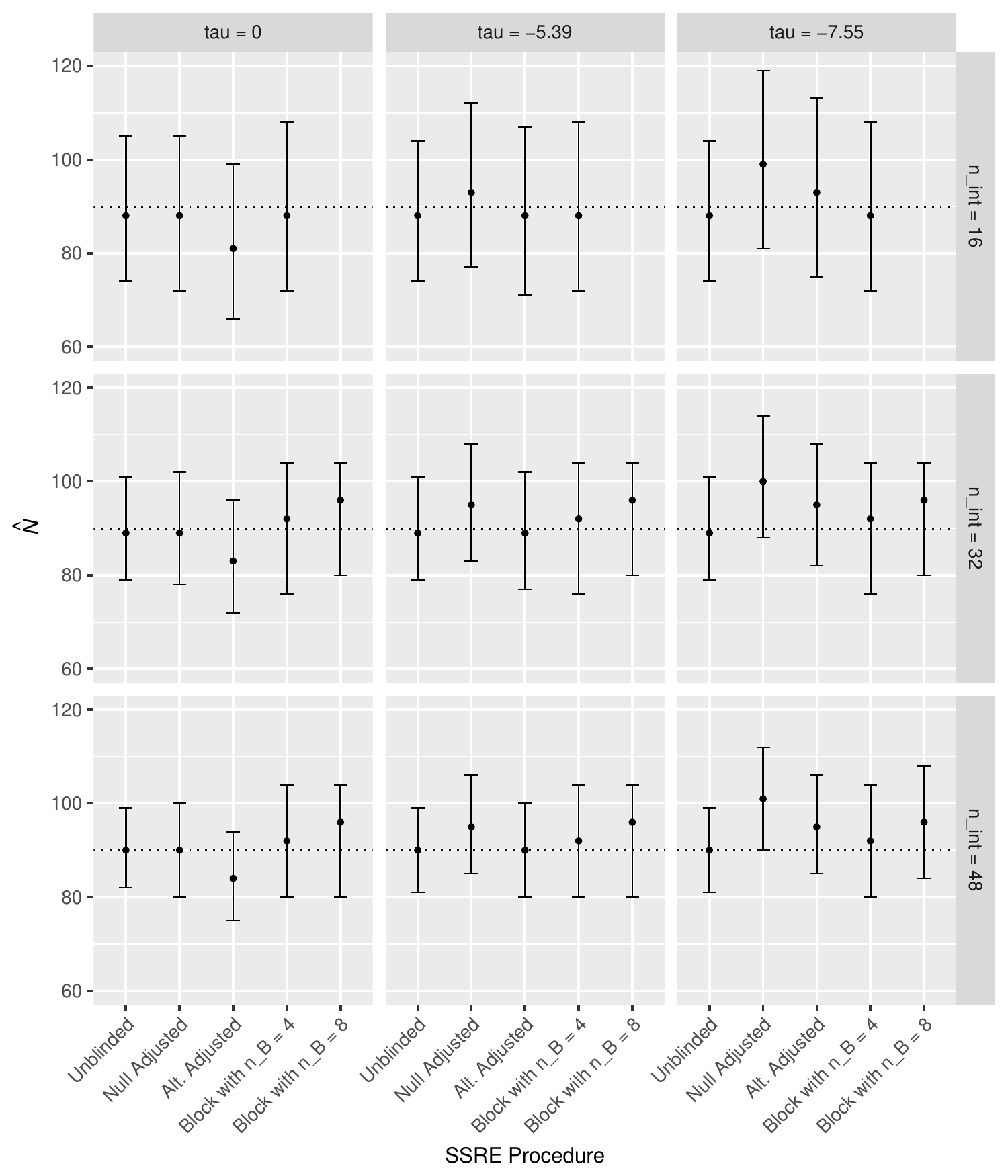}
				\caption{The distribution of $\hat{N}$ is shown for each of the re-estimation procedures for several values of $\boldsymbol{\tau}$, and several values of $n_{\text{int}}$, for Example 3. Precisely, for each scenario, the median, lower and upper quartile values of $\hat{N}$ across the simulations are given. The dashed line indicates the true required value of $N$.}
			\end{center}
		\end{figure}
		
		\subsection{Familywise error-rate and power}
		
		For the scenarios from Section S.M.6.2 that were not conducted under the observed treatment effect, the estimated FWER and power were also recorded. The results are displayed in Table 3.
		
		In this instance, we observe that the FWER is in many instances slightly below the nominal level, with a maximal value of only 0.0252 for the block randomised procedure with $n_B=4$ when $n_{\text{int}}\in\{16,48\}$. This is likely a result of the fact that the number of observations accrued in each of the designs is, for a crossover trial at least, large relative to the number of hypotheses that are to be tested.
		
		Most of the re-estimation procedures attain a power close to the nominal level. However, the alternative adjusted estimator again leads to lower levels of power. Finally, as would be anticipated, increasing the value of $n_{\text{int}}$ improves the power of the re-estimation procedures. However, it has no clear effect upon the FWER.
		
		\subsection{Influence of $\sigma_e^2$}
		
		We now consider cases where $\sigma_b^2=255.0$, and $\delta=-5.39$, but $\sigma_e^2\neq169.8$. Specifically, in Figures 24 and 25 we respectively examine the FWER and power of the re-estimation procedures under the null and alternative hypotheses when $\sigma_e^2\in[0.25(169.8),4(169.8)]$.
		
		Here, there appears to be some evidence that the smallest considered values of $\sigma_e^2$ result in lower values of the FWER. This may once more reflect a change between terminating the trial at the interim reassessment and continuing to the end of the second stage. Nonetheless, is it clear that in this case the procedures control the FWER well regardless of the value of the within person variance.
		
		Similarly, whilst the smallest values of $\sigma_e^2$ lead to the designs being substantially over-powered, it is evident that when $\sigma_e^2$ is increased they are still able to provide approximately the desired power.
		
		\begin{figure}[htb]
			\begin{center}\label{TDS3_sigma_e_FWER}
				\includegraphics[width = 15cm]{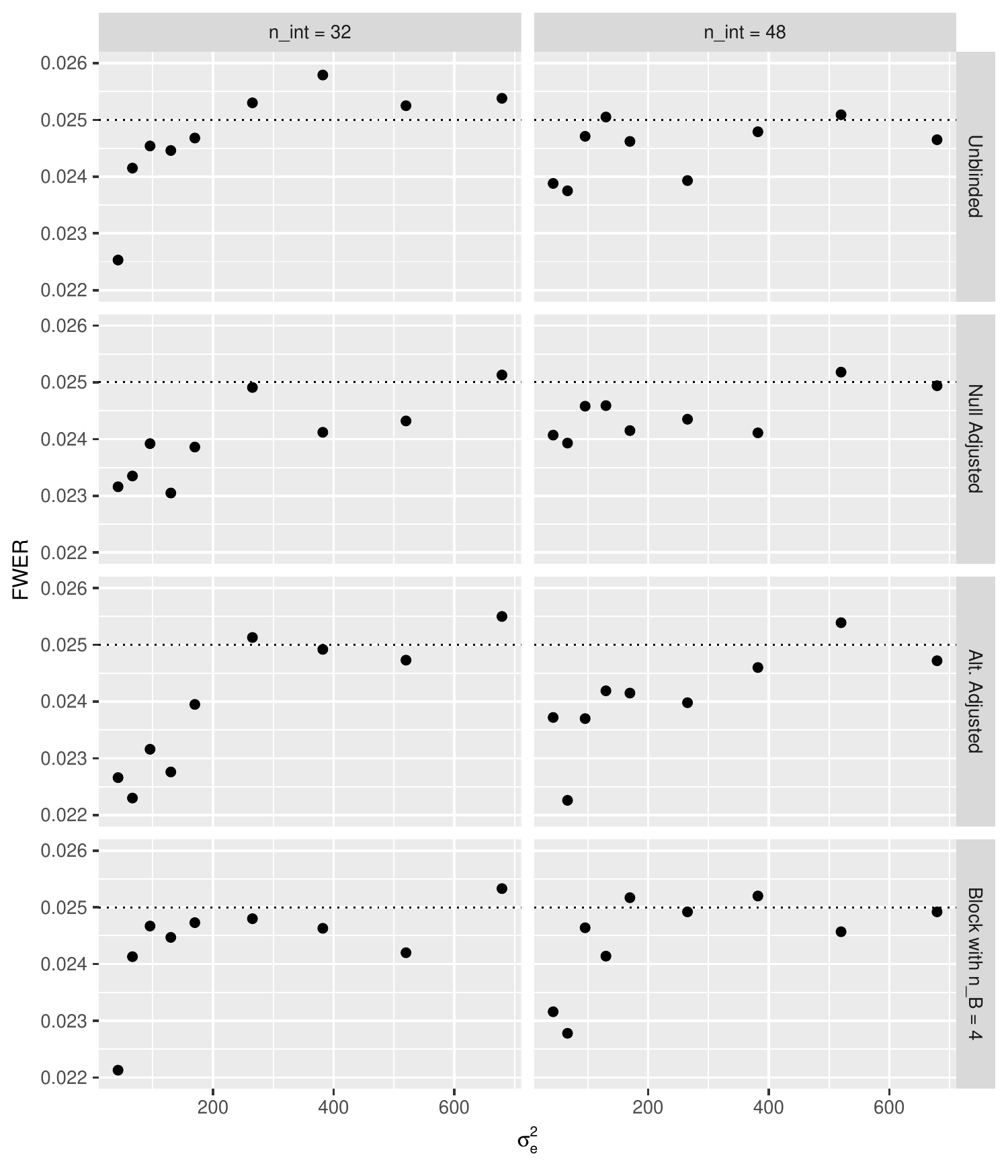}
				\caption{The simulated familywise error-rate (FWER) is shown under the global null hypothesis for each of the re-estimation procedures when $n_{\text{int}}\in\{32,48\}$, as a function of the within person variance $\sigma_e^2$, for Example 1. The Monte Carlo error is approximately 0.0005 in each instance. The dashed line indicates the desired value of the FWER.}		
			\end{center}
		\end{figure}
		
		\begin{figure}[htb]
			\begin{center}\label{TDS3_sigma_e_power}
				\includegraphics[width = 15cm]{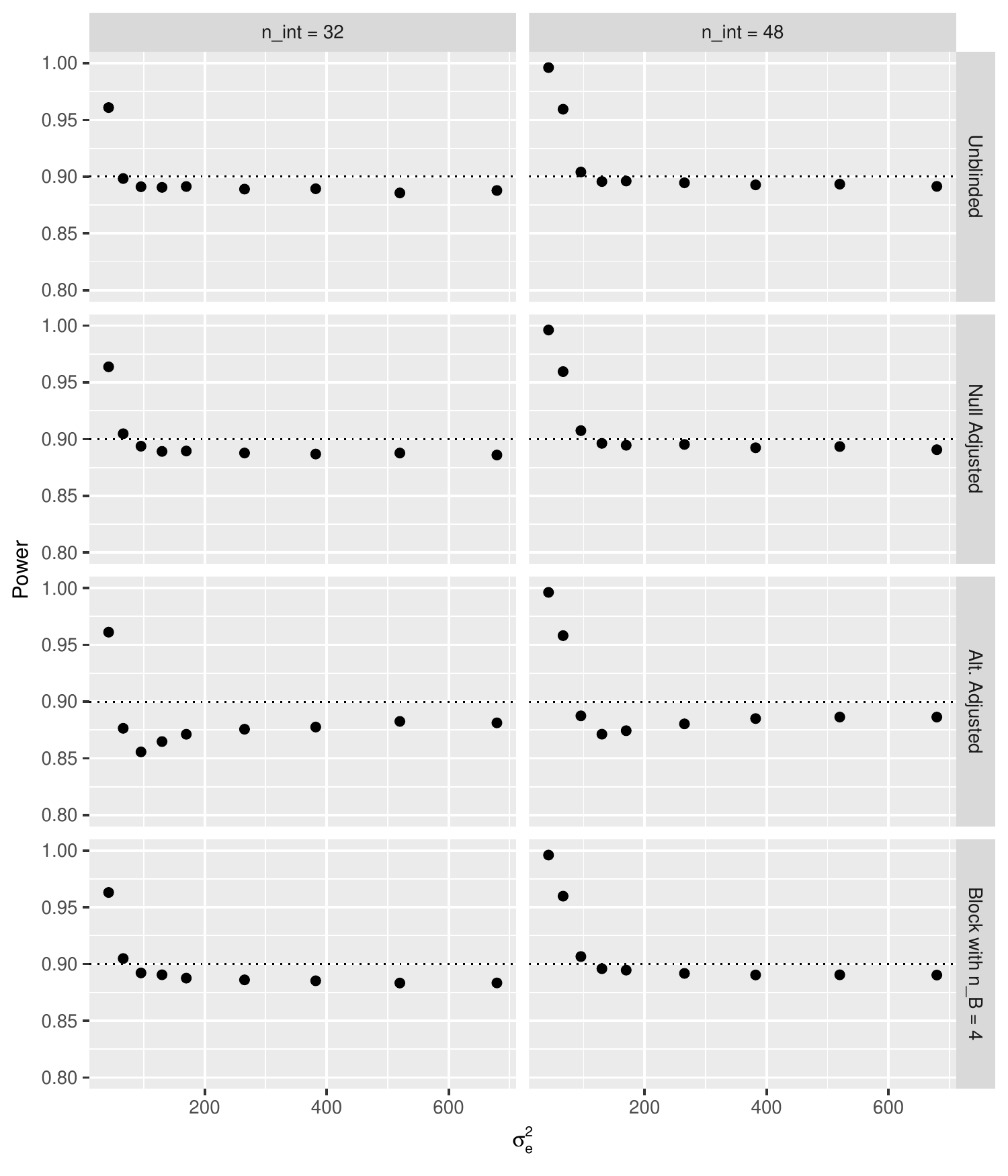}
				\caption{The simulated power is shown under the global alternative hypothesis for each of the re-estimation procedures when $n_{\text{int}}\in\{32,48\}$, as a function of the within person variance $\sigma_e^2$, for Example 1. The Monte Carlo error is approximately 0.001 in each instance. The dashed line indicates the desired value of the power.}
			\end{center}
		\end{figure}
		
		\subsection{Influence of $\sigma_b^2$}
		
		Next, we examine scenarios in which $\sigma_e^2=169.8$, and $\delta=-5.39$, but $\sigma_b^2\neq255.0$: in Figures 26 and 27 we respectively examine the FWER and power of the re-estimation procedures under the null and alternative hypotheses when $\sigma_b^2\in[0.25(255.0),4(255.0)]$.
		
		Similar to both Examples 1 and 2, allowing for Monte Carlo error, the value of $\sigma_b^2$ seems to have little effect upon the ability of the re-estimation procedures to control the FWER to 0.025, and to provide power at level 0.9. We therefore have further evidence that we do not need to be concerned about the underlying between patient variance when considering the appropriate use of the proposed methods.
		
		\begin{figure}[htb]
			\begin{center}\label{TDS3_sigma_b_FWER}
				\includegraphics[width = 15cm]{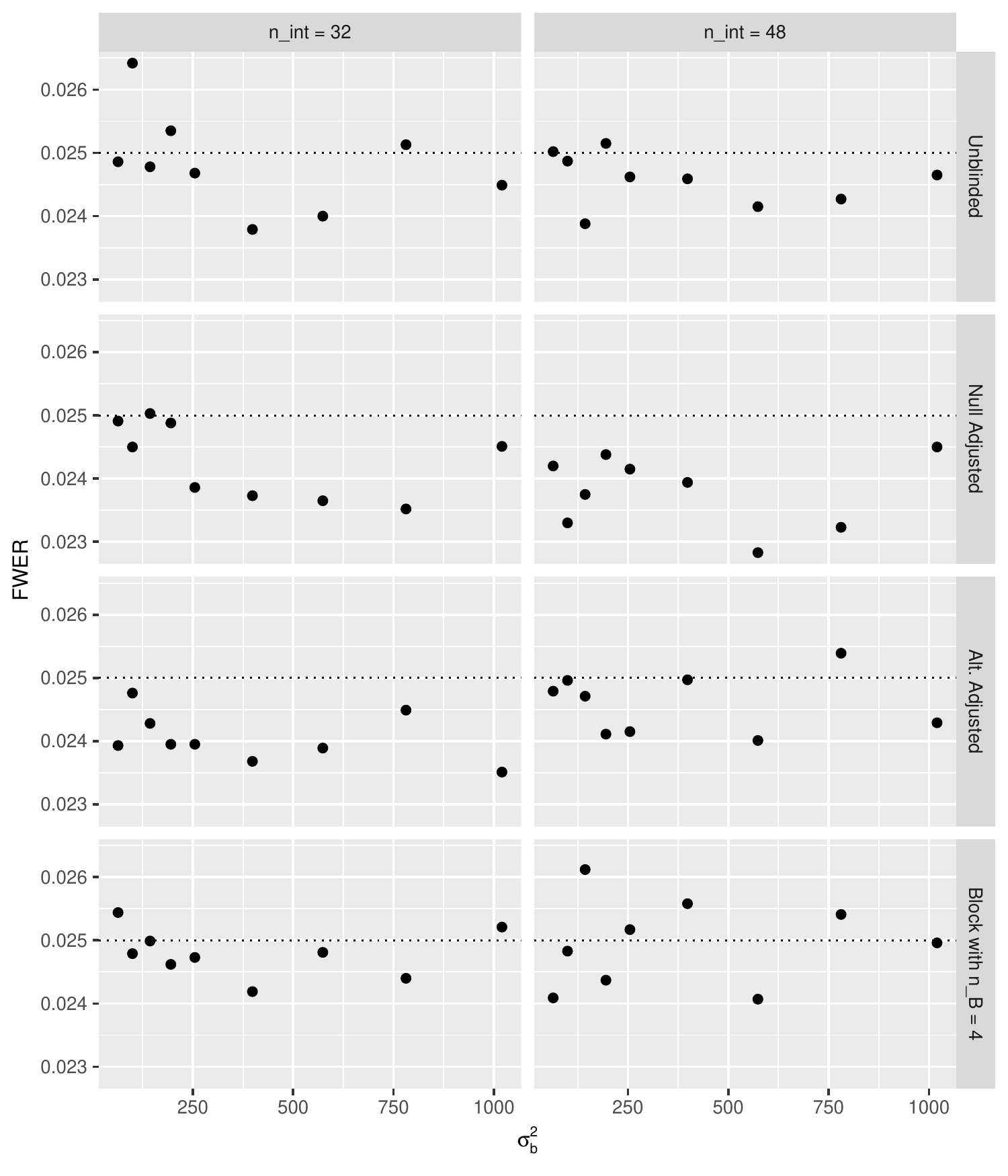}
				\caption{The simulated familywise error-rate (FWER) is shown under the global null hypothesis for each of the re-estimation procedures when $n_{\text{int}}\in\{32,48\}$, as a function of the within person variance $\sigma_b^2$, for Example 3. The Monte Carlo error is approximately 0.0005 in each instance. The dashed line indicates the desired value of the FWER.}		
			\end{center}
		\end{figure}
		
		\begin{figure}[htb]
			\begin{center}\label{TDS3_sigma_b_power}
				\includegraphics[width = 15cm]{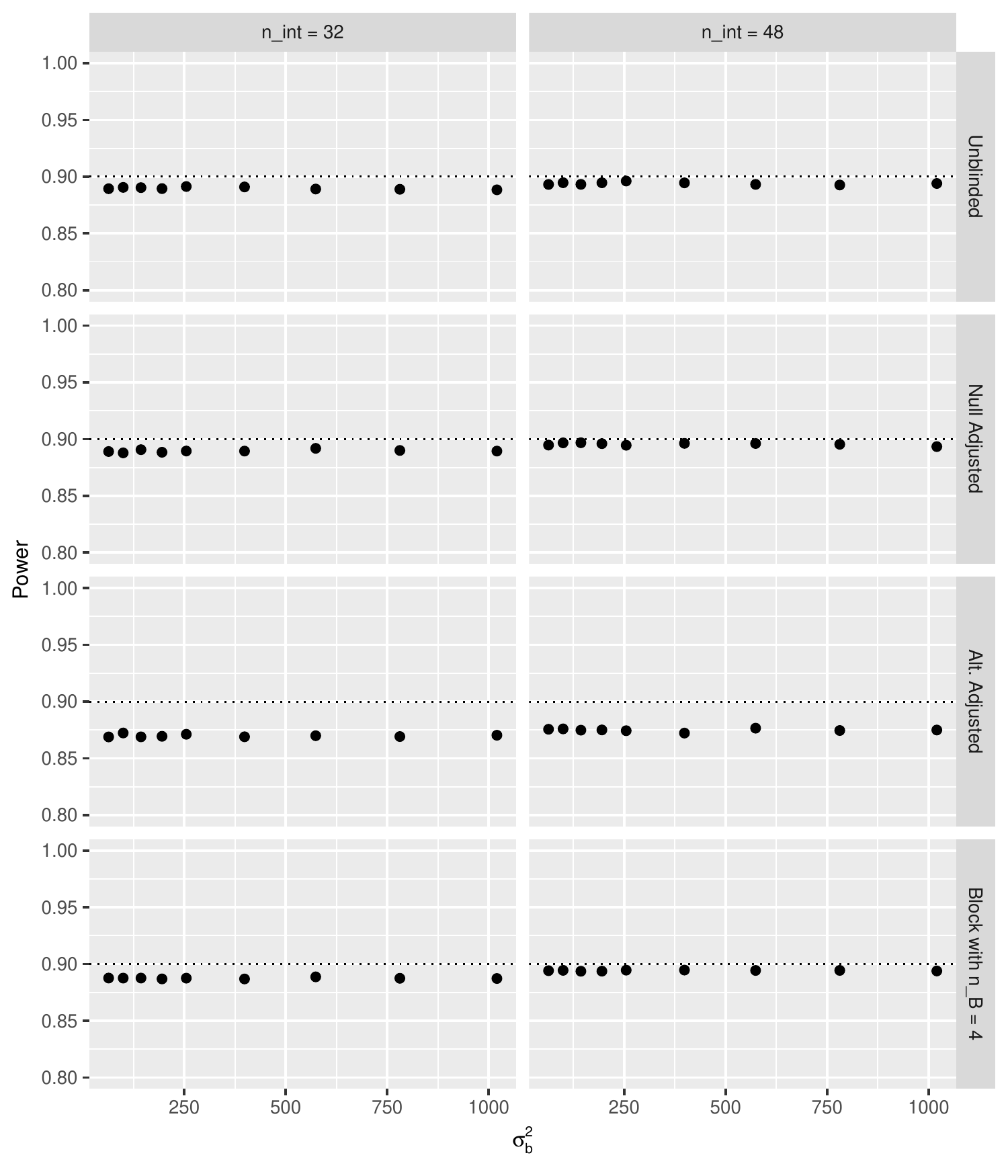}
				\caption{The simulated power is shown under the global alternative hypothesis for each of the re-estimation procedures when $n_{\text{int}}\in\{32,48\}$, as a function of the within person variance $\sigma_b^2$, for Example 3. The Monte Carlo error is approximately 0.001 in each instance. The dashed line indicates the desired value of the power.}
			\end{center}
		\end{figure}
		
		\subsection{Influence of $\delta$}
		
		We now investigate settings in which $\sigma_e^2=169.8$ and $\sigma_b^2=255.0$, but $\delta\neq-5.39$. Precisely, in Figures 28 and 29 we respectively examine the FWER and power of the re-estimation procedures under the null and alternative hypotheses when $\delta\in[2(-5.39),0.5(-5.39)]$.
		
		Our results indicate that in this case, the smallest considered values of $\delta$ appear to result in reduced values of the FWER. However, regardless of the value of $\delta$, each of the procedures is generally able to control the FWER to approximately the nominal level.
		
		From Figure 29, we once more observe that when the magnitude of $\delta$ is large, the designs are over-powered, but outside of this region the specified clinically relevant difference appears to have little effect upon the power.
		
		\begin{figure}[htb]
			\begin{center}\label{TDS3_delta_null}
				\includegraphics[width = 15cm]{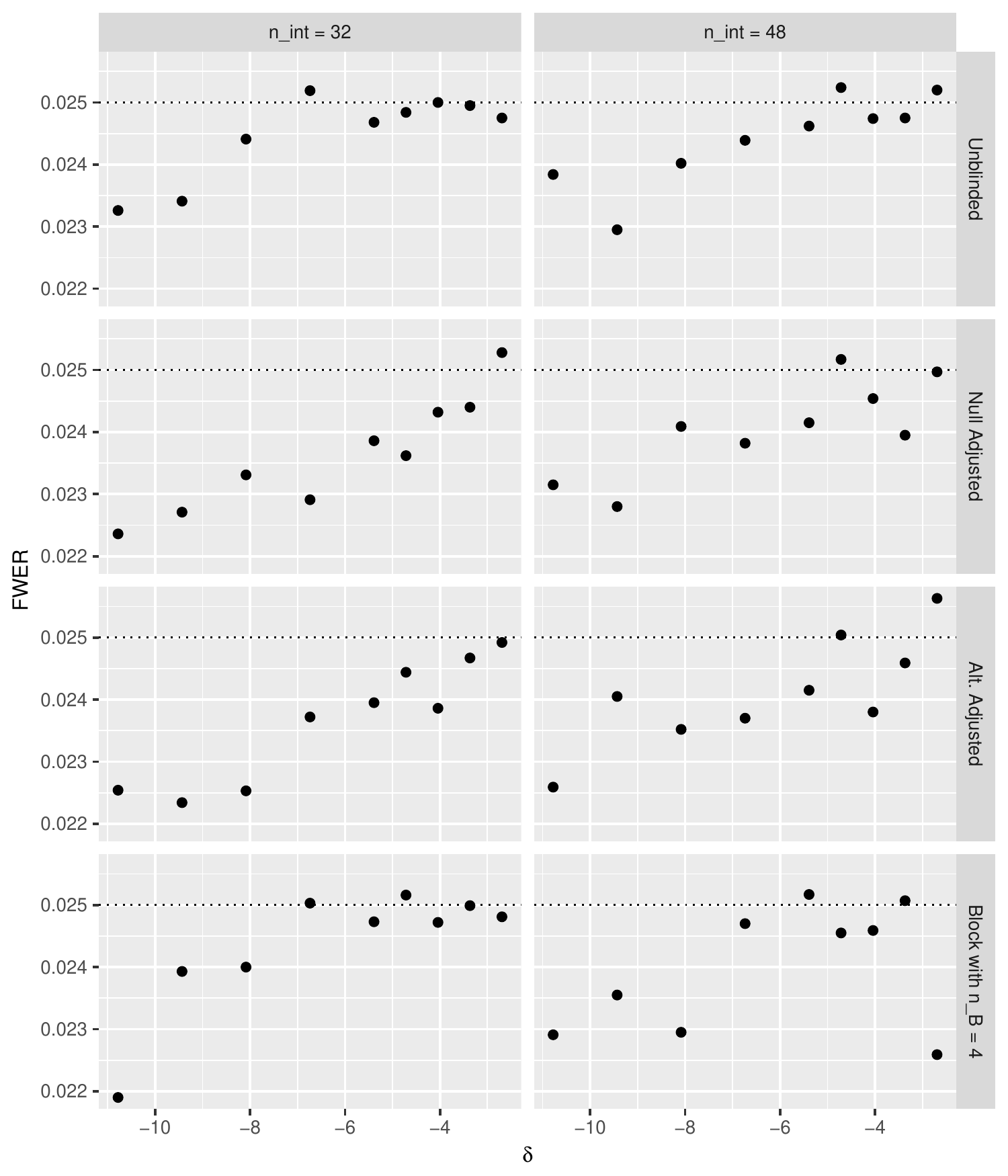}
				\caption{The simulated familywise error-rate (FWER) is shown under the global null hypothesis for each of the re-estimation procedures when $n_{\text{int}}\in\{32,48\}$, as a function of the clinically relevant difference $\delta$, for Example 3. The Monte Carlo error is approximately 0.0005 in each instance. The dashed line indicates the desired value of the FWER.}		
			\end{center}
		\end{figure}
		
		\begin{figure}[htb]
			\begin{center}\label{TDS3_delta_alt}
				\includegraphics[width = 15cm]{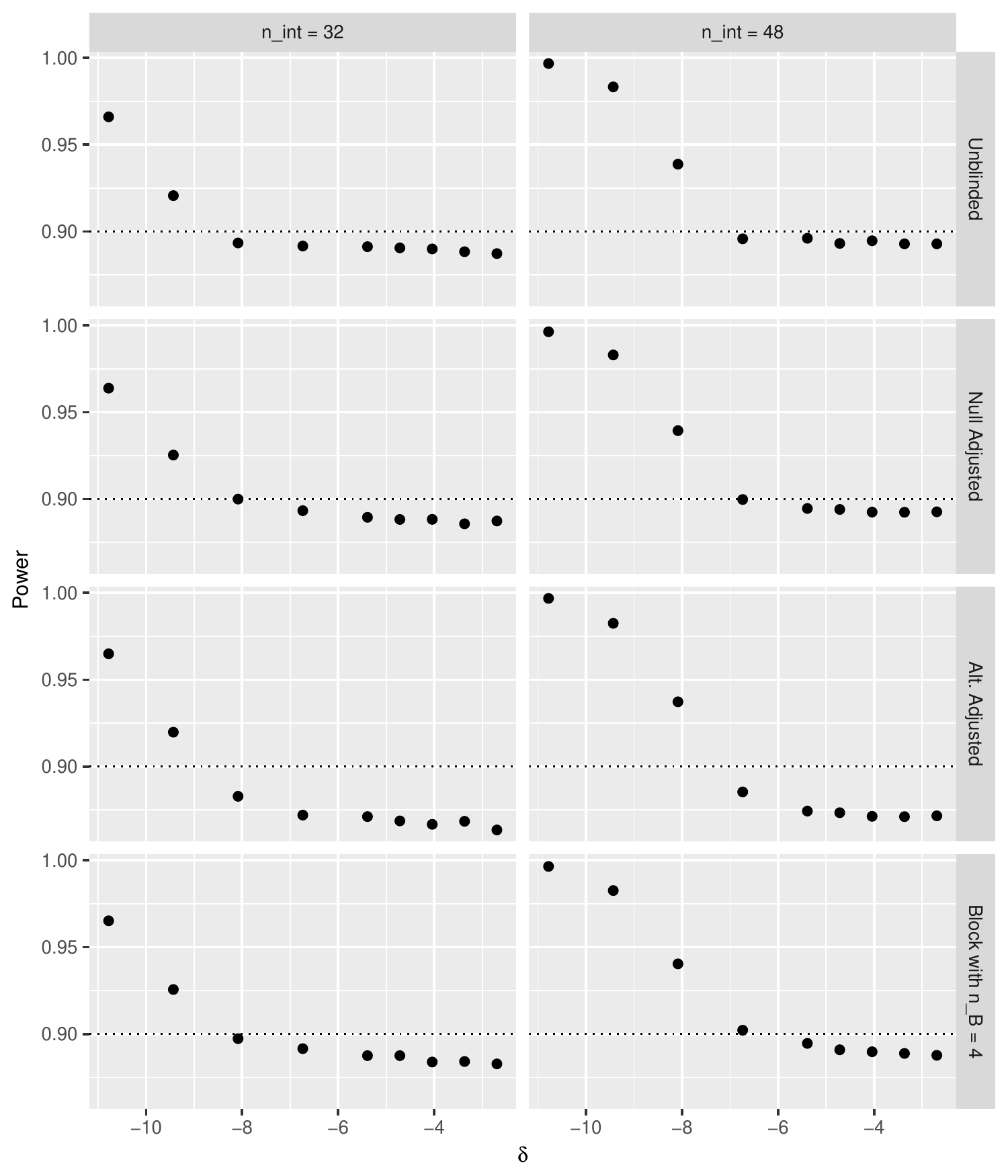}
				\caption{The simulated power is shown under the global alternative hypothesis for each of the re-estimation procedures when $n_{\text{int}}\in\{32,48\}$, as a function of the clinically relevant difference $\delta$, for Example 3. The Monte Carlo error is approximately 0.001 in each instance. The dashed line indicates the desired value of the power.}
			\end{center}
		\end{figure}
		
		\subsection{Sample size inflation factor}
		
		Our final consideration is to once more examine the utility of the sample size inflation factor introduced in Section 3.6. In Figure 30 we present the power of the re-estimation procedures under the alternative hypothesis, when $n_{\text{int}}\in\{16,32,48\}$, with and without the use of the inflation factor.
		
		As was observed for Example 1, the inflation factor leads to a notable boost in power, helping several of the estimators achieve the desired performance. Therefore, when there are concerns around attaining the desired power, we may again recommend that investigators consider utilising this simple adjustment.
		
		\begin{figure}[htb]
			\begin{center}\label{TDS3_inflationfactor}
				\includegraphics{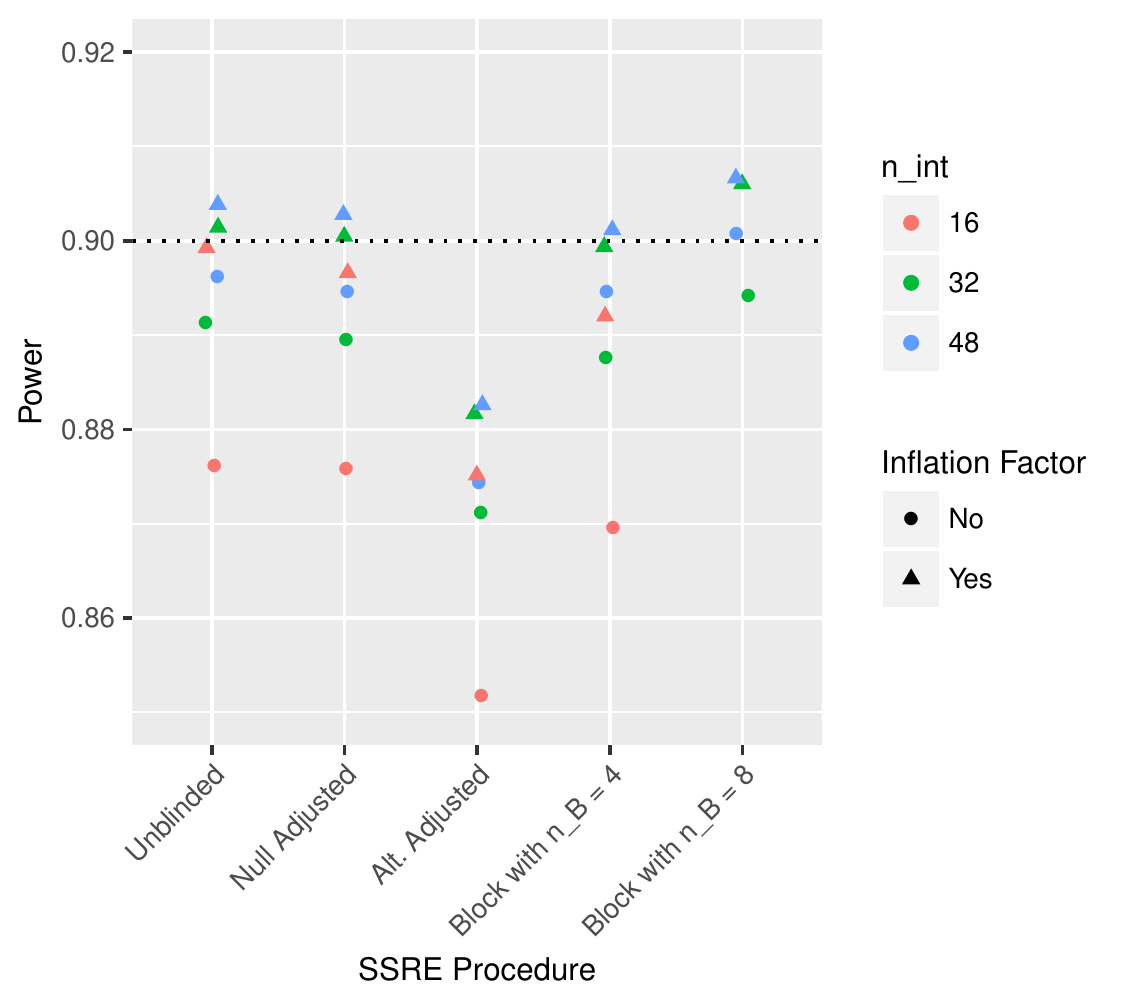}
				\caption{The influence of the considered inflation factor upon the power of the re-estimation procedures under the global alternative hypothesis is shown for several values of $n_{\text{int}}$, for Example 3. The dashed line indicates the desired value of the power.}
			\end{center}
		\end{figure}

%\newpage
%\phantom{aaaa}

\begin{table}[htb]
	\label{tab1}
	\begin{center}
		\caption{The estimated familywise error-rate (FWER) is shown for each of the considered re-estimation procedures and several values of $n_{\text{int}}$ under the global null hypothesis, for Example 1. Corresponding values of the power when only treatment one is effective, treatments one and two are effective, or under the global alternative hypothesis when all three experimental treatments are effective, are also shown. The Monte Carlo error of the FWER and power values is approximately 0.0007 and 0.0013 respectively in each instance. All figures are given to four decimal places.}
		\begin{tabular}{lrrrrr}
			\hline
			& & & \multicolumn{3}{c}{Power} \\
			Re-estimation procedure  & $n_{\text{int}}$ & FWER & $\boldsymbol{\tau} = (\delta,0,0)$ & $\boldsymbol{\tau} = (\delta,\delta,0)$ & $\boldsymbol{\tau} = (\delta,\delta,\delta)$ \\
			\hline
			Unblinded                  & 8  & 0.0513 & 0.7704 & 0.7694 & 0.7687\\
			Null Adjusted              & 8  & 0.0496 & 0.7743 & 0.7809 & 0.7753\\
			Alt. Adjusted              & 8  & 0.0500 & 0.7440 & 0.7512 & 0.7432\\
			Block rand. with $n_B = 2$ & 8  & 0.0509 & 0.7443 & 0.7455 & 0.7428\\
			Unblinded                  & 16 & 0.0506 & 0.7906 & 0.7893 & 0.7867\\
			Null Adjusted              & 16 & 0.0512 & 0.7956 & 0.8010 & 0.7942\\
			Alt. Adjusted              & 16 & 0.0495 & 0.7702 & 0.7731 & 0.7691\\
			Block rand. with $n_B = 2$ & 16 & 0.0512 & 0.7720 & 0.7723 & 0.7747\\
			Block rand. with $n_B = 4$ & 16 & 0.0525 & 0.7858 & 0.7887 & 0.7868\\
			Unblinded                  & 24 & 0.0509 & 0.7963 & 0.7934 & 0.7950\\
			Null Adjusted              & 24 & 0.0496 & 0.8019 & 0.8071 & 0.7990\\
			Alt. Adjusted              & 24 & 0.0508 & 0.7776 & 0.7793 & 0.7770\\
			Block rand. with $n_B = 2$ & 24 & 0.0504 & 0.7821 & 0.7838 & 0.7835\\
			Unblinded                  & 32 & 0.0520 & 0.7977 & 0.7962 & 0.7988\\
			Null Adjusted              & 32 & 0.0509 & 0.8055 & 0.8109 & 0.8072\\
			Alt. Adjusted              & 32 & 0.0498 & 0.7772 & 0.7857 & 0.7812\\
			Block rand. with $n_B = 2$ & 32 & 0.0514 & 0.7907 & 0.7879 & 0.7887\\
			Block rand. with $n_B = 4$ & 32 & 0.0511 & 0.8014 & 0.8002 & 0.8035\\
			Unblinded                  & 40 & 0.0516 & 0.7967 & 0.8010 & 0.8000\\
			Null Adjusted              & 40 & 0.0504 & 0.8081 & 0.8115 & 0.8062\\
			Alt. Adjusted              & 40 & 0.0498 & 0.7828 & 0.7858 & 0.7842\\
			Block rand. with $n_B = 2$ & 40 & 0.0518 & 0.7914 & 0.7926 & 0.7942\\
			\hline
		\end{tabular}
	\end{center}
\end{table}

\begin{table}[htb]
	\label{tab2}
	\begin{center}
		\caption{The estimated familywise error-rate (FWER) is shown for each of the considered re-estimation procedures under the global null hypothesis, for Example 2. Corresponding values of the power when only treatment one is effective, under the global alternative hypothesis when both experimental treatments are effective, and under the global alternative hypothesis with use of the sample size inflation factor, are also shown. The Monte Carlo error of the FWER and power values is approximately 0.001 and 0.0013 respectively in each instance. All figures are given to four decimal places.}
		\begin{tabular}{lrrrr}
			\hline
			& & \multicolumn{3}{c}{Power} \\
			Re-estimation procedure  & FWER & $\boldsymbol{\tau} = (\delta,0)$ & $\boldsymbol{\tau} = (\delta,\delta)$ & $\boldsymbol{\tau} = (\delta,\delta)$ with Inf. Fac. \\
			\hline
			Unblinded                  & 0.1174 & 0.8027 & 0.8047 & 0.8276 \\
			Null Adjusted              & 0.1069 & 0.8204 & 0.8186 & 0.8450 \\
			Alt. Adjusted              & 0.1069 & 0.7490 & 0.7485 & 0.7710 \\
			Block rand. with $n_B = 3$ & 0.1157 & 0.8013 & 0.8029 & 0.8273 \\
			\hline
		\end{tabular}
	\end{center}
\end{table}

\begin{table}[htb]
	\label{tab3}
	\begin{center}
		\caption{The estimated familywise error-rate (FWER) and power is shown for each of the considered re-estimation procedures and several values of $n_{\text{int}}$ under the null and alternative hypotheses respectively, for Example 3. The Monte Carlo error of the FWER and power values is approximately 0.0005 and 0.001 respectively in each instance. All figures are given to four decimal places.}
		\begin{tabular}{lrrr}
			\hline
			Re-estimation procedure    & $n_{\text{int}}$ & FWER & Power \\
			\hline
			Unblinded                  & 16 & 0.0243 & 0.8761 \\
			Null Adjusted              & 16 & 0.0243 & 0.8758 \\
			Alt. Adjusted              & 16 & 0.0237 & 0.8517 \\
			Block rand. with $n_B = 4$ & 16 & 0.0252 & 0.8696 \\
			Unblinded                  & 32 & 0.0247 & 0.8913 \\
			Null Adjusted              & 32 & 0.0239 & 0.8895 \\
			Alt. Adjusted              & 32 & 0.0240 & 0.8712 \\
			Block rand. with $n_B = 4$ & 32 & 0.0247 & 0.8876 \\
			Block rand. with $n_B = 8$ & 32 & 0.0241 & 0.8942 \\
			Unblinded                  & 48 & 0.0246 & 0.8961 \\
			Null Adjusted              & 48 & 0.0242 & 0.8946 \\
			Alt. Adjusted              & 48 & 0.0242 & 0.8744 \\
			Block rand. with $n_B = 4$ & 48 & 0.0252 & 0.8946 \\
			Block rand. with $n_B = 8$ & 48 & 0.0249 & 0.9007 \\
			\hline
		\end{tabular}
	\end{center}
\end{table}

\end{document}